\documentclass[a4paper,fleqn]{cas-dc}
\usepackage[numbers]{natbib}
\usepackage{caption}
\usepackage{url}
\usepackage{xcolor}
\usepackage{color,soul}
\usepackage[textwidth=  15mm]{todonotes}
\usepackage{environ} 
\usepackage{etoolbox}
\usepackage{lipsum} 
\usepackage{blindtext}
\usepackage[title]{appendix}

\graphicspath{{./images/}}

\newtoggle{doShowFigures}        
\NewEnviron{maybeShowFigures}{%
  \iftoggle{doShowFigures}{\BODY}{}%
}
\toggletrue{doShowFigures}
\begin{document}
\let\WriteBookmarks\relax
\def\floatpagepagefraction{1}
\def\textpagefraction{.001}

\shorttitle{Majority Voting Approach to Ransomware Detection}
\shortauthors{S.R.Davies,R.Macfarlane,W.J.Buchanan}

\title [mode = title]{Majority Voting Approach to Ransomware Detection}                      
\author[1]{Simon R. Davies}[type=editor,
                        auid=000,bioid=1,
                        orcid=0000-0001-9377-4539,]
\cormark[1]
\ead{s.davies@napier.ac.uk}
\address[1]{School of Computing, Edinburgh Napier University, Edinburgh, UK.}

\author[1]{Richard Macfarlane}[orcid=0000-0002-5325-2872]

\author[1]{William J. Buchanan}[orcid=0000-0003-0809-3523]

\begin{abstract}
Crypto-ransomware remains a significant threat to governments and companies alike, with high-profile cyber security incidents regularly making headlines. Many different detection systems have been proposed as solutions to the ever-changing dynamic landscape of ransomware detection. In the majority of cases, these described systems propose a method based on the result of a single test performed on either the executable code, the process under investigation, its behaviour, or its output. In a small subset of ransomware detection systems, the concept of a scorecard is employed where multiple tests are performed on various aspects of a process under investigation and their results are then analysed using machine learning. 

The purpose of this paper is to propose a new majority voting approach to ransomware detection by developing a method that uses a cumulative score derived from discrete tests based on calculations using algorithmic rather than heuristic techniques. The paper describes 23 candidate tests, as well as 9 Windows API tests which are validated to determine both their accuracy and viability for use within a ransomware detection system. 

Using a cumulative score calculation approach to ransomware detection has several benefits, such as the immunity to the occasional inaccuracy of individual tests when making its final classification. The system can also leverage multiple tests that can be both comprehensive and complimentary in an attempt to achieve a broader, deeper, and more robust analysis of the program under investigation. Additionally, the use of multiple collaborative tests also significantly hinders ransomware from masking or modifying its behaviour in an attempt to bypass detection.

The results achieved by this research demonstrate that many of the proposed tests achieved a high degree of accuracy in differentiating between benign and malicious targets and  suggestions are offered as to how these tests, and combinations of tests, could be adapted to further improve the detection accuracy.
\end{abstract}

\begin{keywords}
\sep Ransomware Detection \sep Malice Score \sep Score card \sep Malware \sep
\sep
\sep
----------------------------------------------
\sep
Article history
\sep
Received 1\textsuperscript{st} April 2023
\sep
Accepted ?
\sep
Available
\end{keywords}

\maketitle
\section{Introduction}
\label{cha:intro}
\noindent Crypto-ransomware infections remain a significant threat to governments and companies alike with high-profile cyber security incidents regularly making headlines~\cite{ION,royalmail}. The detection of ransomware is often described as an arms race~\cite{Oz2021} between threat actors and the people responsible for developing effective malware countermeasures and techniques.

There are two main approaches used in malware analysis in general and ransomware analysis in particular. Static Analysis~\cite{Yamany2022}, where the evaluation of the program is performed without the actual execution of the code. Essentially the program contents are examined in an attempt to determine the nature of the program and its possible application. This is normally achieved by attempting to isolate and identify known patterns or signatures within the code. Static analysis scales well and can provide better coverage of a ransomware binary code. However, static analysis can produce false execution behaviour as code paths may not be reachable during actual execution~\cite{Dutta2022} and tell-tale signatures may not be known at the time of analysis.
Dynamic Analysis, on the other hand, executes the program under investigation in an instrumented or monitored manner and garners more factual information on the behaviour and effect of the program. Dynamic analysis can provide more accurate information on the actual execution behaviour of the investigated binary, though dynamic analysis can be computationally expensive~\cite{lebbie2022} and contains some element of risk.

The problem of automatic malware detection is a difficult one, with no full solution in sight despite decades of  research~\cite{DeGaspari2020a}. The traditional approach, based on analysis of static signatures - is increasingly rendered ineffective by polymorphism and the widespread availability of program obfuscation tools~\cite{moser,kane}. Using such tools, malware creators can quickly generate thousands of binary variants of functionally identical samples, effectively circumventing \\ signature-based approaches.
As a result, in recent years, the focus of the research community has increasingly shifted toward dynamic, behaviour-based analysis techniques. Behavioural approaches \newline sidestep the challenges of obfuscated binary analysis. Instead, they focus on the run-time behaviour of malware processes, which is difficult to alter without breaking core functionality and is therefore considered a reliable fingerprint for malware presence~\cite{DeGaspari2020a}.

Over the years many different detection systems have been proposed as solutions to the ever-changing dynamic landscape of ransomware detection. These approaches have leveraged many different techniques such as machine learning \cite{Yamany2022,Ahmed2021,Ahmed2020a,Kim2022,Salehi2018,Scaife2016}, neural networks \cite{Alam2020,Homayoun2020,Lokuketagoda2018,McDonald2022}, file entropy \cite{Scaife2016,Hall2006,Lee2019a,Lee2019,VandenBrink2016}, kernel hooking and process behaviour \cite{Al-rimy2019,Bottazzi2018,Ki2015,Song2016}. In the majority of cases, the described systems propose a method based on the result of a single test performed on either the executable code, the process under investigation, its behaviour or its output. Many of the proposed systems claim to archive relatively high accuracy. Unfortunately, the researchers rarely publish enough detail of their research or the datasets used to allow the reported results to be replicated. Berrueta~\cite{Berrueta2019} identifies that there are no common metrics of accuracy and performance in ransomware detection. The fragmentation of scientific research on ransomware combined with a lack of coherent investigation methodology is a major challenge in this research~\cite{Dargahi2019}. This view is supported by Maigida~\cite{Maigida2019} who state that the lack of readily available data is also hindering the speedy development of detection and prevention solutions.

In a small subset of ransomware detection systems, the concept of a scorecard is employed. In these specific detection systems, multiple tests are performed on various aspects of a process under investigation. The results of each test contribute to an overall score for the process. A decision can then be made, based on this score, as to whether the process under investigation is benign or malicious. The main proponent of this approach was Kharraz~\cite{Kharraz2017a} in their implementation of the \emph{Redemption} detection system. In this work, they refer to this cumulative score as a~\emph{Malice Score}, and for the remainder of this paper, we will use their terminology when discussing this combined ranking score. Other detection systems that have also used this concept of a cumulative malice score are~\cite{abbasi,Continella,john,Kharaz2016,Mehnaz2018}. 

None of the described systems used an analytical or algorithmic approach to calculating values that could then be combined into a cumulative malice score, rather they relied on some form of machine learning to determine the result. This paper describes the work performed by the authors in building on the original research conducted by Kharraz~\cite{Kharraz2017a}, enhancing and updating their approach and proposing many new discretely calculated  static and dynamic analysis tests that could be incorporated into the final malice score calculation. 

A majority voting approach was chosen for the ransomware detection system proposed in this work. With this type of system, each of the underlying contributing tests generates a binary output. The result of an individual test can be either that it is considered malicious or it can be considered benign. These individual contributing scores are calculated using algorithmic rather than the heuristic techniques previously proposed in earlier research. Once all the tests have been performed, the resulting votes are then collated into two sets, malicious votes and benign votes. The final classification decision of the detection system is then determined from the set that received the majority of votes. An advantage of this approach is that the system requires no training, as the constituent values are calculated using discrete reproducible tests that require no prior knowledge or model training. These proposed new additional tests are validated using a modern and diverse dataset~\cite{Davies2022} to determine both their accuracy and viability for use within a ransomware detection system. In the initial design, each test has an equal weighting and thus an equal contribution to the final result. However, this design may be adapted in later iterations by the inclusion of weighting and bias to the results of individual tests, allowing their votes to have more effect on the final decision. 

There are many benefits associated with using a cumulative score calculation approach to ransomware detection. For example, when using such an approach, the detection system does not rely on a single specific attribute to base its decision on whether the program under investigation is malicious or not. Rather it can leverage multiple tests that can be both comprehensive and complimentary in an attempt to achieve a broader, deeper and more robust analysis of the program under investigation. Also, such a system would be easier to enhance, as adding additional tests based on new research would be straightforward. Bias from one particular test~\cite{Yamany2022,DeGaspari2020a} would also be mitigated, and the weighting of each contributing test could be adjusted to improve accuracy. Additionally, the use of multiple collaborative tests also significantly hinders ransomware from masking or modifying its behaviour in its attempt to bypass detection~\cite{DeGaspari2020a,mcintosh}.

The remainder of the paper is structured as follows. In Section 2, we discuss some of the main techniques used in ransomware detection and discuss in detail other techniques that use a collaborative voting approach or a combined scoring technique. In Section 3, we provide a description of the candidate tests that could potentially be included in the cumulative malice scoring calculation and outline the methodology used in the experiments. In Section 4, we present the recorded results and discuss the consequences of the findings with regard to the development of anti-ransomware techniques, and we provide some recommendations for crypto-ransomware detection approaches moving forward. Finally, in Section 5, we discuss the main findings and conclusions gained from this research together with possible limitations in using this approach and suggest further research that could be conducted based on the findings from the research presented in this paper.

\noindent The main contributions of this paper are:
\begin{itemize}
    \item Design, development and detailed description of 23 potential ransomware detection tests.
    \item Investigation into the amount and frequency of Windows API calls within the ransomware executable files and volatile memory of a ransomware process.
    \item Validation of the effectiveness of the proposed tests in detecting ransomware.
    \item A ransomware detection system based on algorithmic derived ransomware indicators.
    \item The use of a modern publicly available dataset during the development and testing of the system. The majority of the similar systems proposed in the literature use datasets that are up to 14 years old.
\end{itemize}

\section{Related Work}
\noindent Over the last 20 years, a significant number of ransomware detection systems have been proposed in the research literature. The approaches used by these detection systems can be loosely divided into two categories. In one approach, a single method or test is developed which is then used to determine if the system is being attacked by ransomware. The alternative approach is to use machine learning to perform the identification. With the machine learning approach, the system designers identify key features from the running process and system under investigation. The machine learning model then attempts to determine patterns within these features on which to base its judgement. A decision, or classification, is then made, based on the measured values of these features, as to whether the system is under attack or not.

Examples of single-method approaches are~\cite{Ahmed2021,Kim2022,Ganfure2020,Manavi2022,Prachi2022,Sheen2022}. In these cases, the entire effectiveness of the detection technique relies solely on the ability of this single criterion to distinguish between benign and malicious programs~\cite{DeGaspari2023}. For example, one particular technique used in the identification of ransomware execution is to use the calculated entropy value of the files created by a process. Encrypted files tend to have a high entropy value whereas the entropy value of plain text files is much lower. Encrypted output files generated during the execution of a ransomware program would tend to have higher entropy values, possibly allowing them to be identified as a product of a ransomware infection. Unfortunately, this technique struggles to correctly distinguish between encrypted files and benign files that also have high entropy such as compressed files. The use of entropy as a detection metric has also been called into question~\cite{mcintosh,lee2022} as there exist techniques that could be used by ransomware to avoid detection via encoding or, in some other way, manipulating the encrypted output file.

Examples of ransomware detection techniques that have leveraged machine learning are~\cite{Yamany2022,Ahmed2021,Ahmed2020a,Kim2022,Salehi2018,Scaife2016} or similarly neural networks~\cite{Alam2020,Homayoun2020,Lokuketagoda2018,McDonald2022}. These systems are trained using extracted features from typical ransomware processes or systems that are being attacked by ransomware. Examples of features that are used in these systems are: write entropy, file overwrite behaviour, directory traversal, directory listing, cross-file type access, read/write/ create/close operations, temporary files, file type coverage, file similarity, file type change and access frequency~\cite{DeGaspari2023}. In most cases, with systems that rely on machine learning to determine if a system is being attacked, the significance of the individual extracted features and their subsequent impact on the final classification is represented internally by the detection system's model and is not immediately obvious to an observer. Inadequacies with this approach have been investigated in the literature~\cite{DeGaspari2023} which discusses classifier evasion techniques, known as adversarial machine learning that can be leveraged by ransomware developers to avoid classification and subsequent detection.

However, in a few proposed ransomware detection systems, the designers do try to provide insight into the machine-learning techniques used and how the tested features affect the overall decision-making process. The developers of the detection system UNVEIL~\cite{Kharaz2016} and its successor Redemption~\cite{Kharraz2017a}, introduce the concept of a~\emph{malice score} which is a combined weighted score derived from the outcome of individual feature tests. The system detects suspicious activity using dynamic analysis and generates a malice score using a heuristic function. Inputs to this function are various behavioural features such as file entropy changes, writes that cover extended portions of a file, file deletion, processes writing to a large number of user files, processes writing to files of different types and back-to-back writes.  CryptoLock~\cite{Scaife2016a} propose a similar approach summing the results of various tests into a cumulative scoring they refer to as a \emph{Reputation Score}. This score is derived from measurements of file type changes, the similarity between original and written content and output file entropy values. Another detection system, RWGuard~\cite{Mehnaz2018}, does mention the specific features that are inspected and include file IO, decoy files, file change monitoring and crypto API monitoring. However, very little detail on how the specific calculations are performed is provided. DNA-Droid~\cite{Gharib2017}, was the only detection system found that, leveraged a combination of static and dynamic analysis as the inputs to their neural network model. In all cases if this cumulative score is above a certain threshold, then the process is deemed to be malicious, otherwise, the process is considered benign. 

However, in all these cases, the individual test results and thresholds are still determined heuristically via the machine-learning model. The model itself decides the significance and weighting given to each extracted feature and the influence that each feature has on the final classification. Reducing the entire decision-making process to effectively a black box function. A consequence of this is that it is difficult for the designers to directly affect the final decision, thus preventing them from being easily able to tune and influence the decision-making process and final classification produced by the model. The resulting quality and accuracy of the decisions made by these systems are essentially reliant on the quality of the training data used to develop the models in the first place.

No ransomware detection systems have been identified in the literature that uses a malice scoring type approach where the constituent scores contributing to the final malice score are determined using analytical or algorithmic calculation methods as opposed to the heuristics used in machine learning approaches.

\section{Methodology}
\label{cha:methodology}
\begin{figure*}
    \begin{center}   
  \includegraphics[scale=0.4]{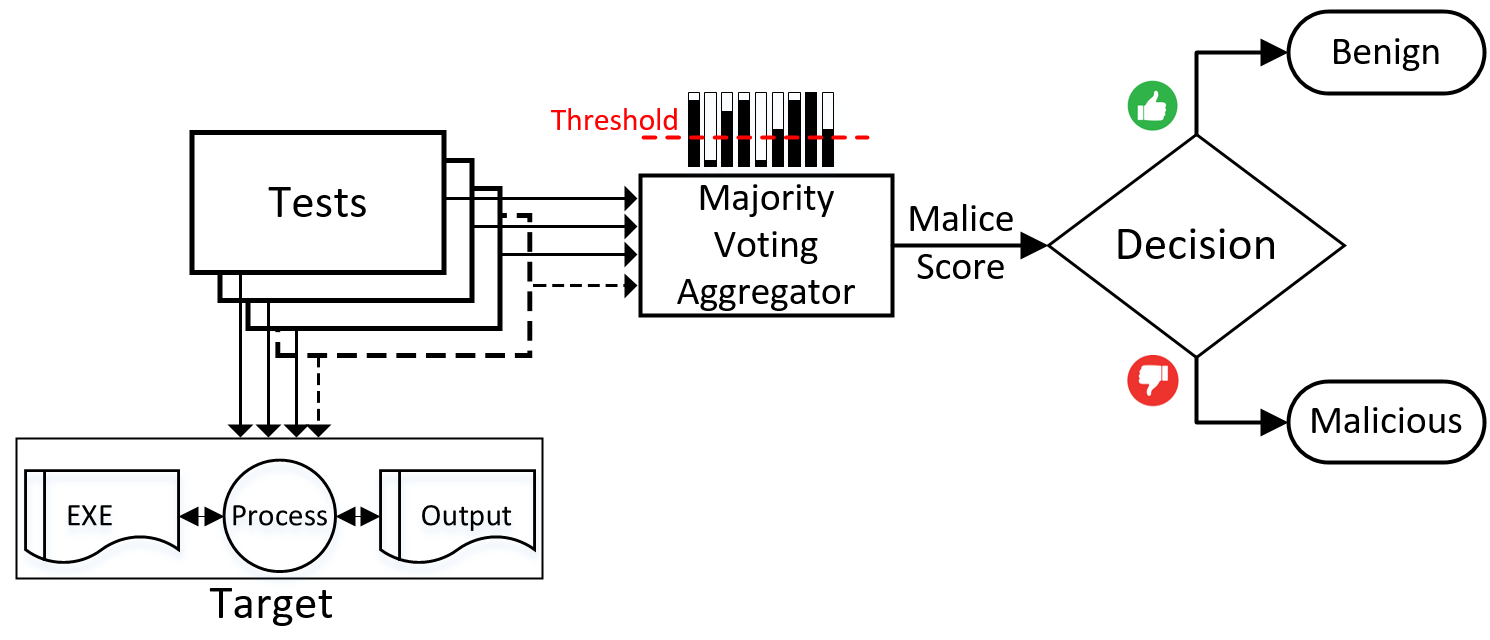}
  \caption{Overview of Proposed System}
  \label{fig:system-overview}
  \end{center}
\end{figure*}

\noindent This section introduces a collection of potential tests that could be used in collaboration to determine if a process is malicious or benign. There is a binary outcome for each of these tests with a test failure indicating that the subject of the test is more likely to be malicious and passing the test indicating that it is more likely to be benign. The resulting votes from each test are then recorded. Each of these proposed test results would then contribute to the final overall \emph{malice score} of the process under investigation. Each contributing test has the same weighting and thus the same impact on the final scoring. After all the tests have been conducted the classification decision is made, based on an aggregation of received  votes, malicious or benign. A conceptual overview of how the proposed system would be configured is shown in Figure~\ref{fig:system-overview}.

\subsection{File Content Analysis}
\label{filecontentanalysis}
\noindent This collection of tests is performed on any output produced by the process. In the majority of cases, this would manifest itself as files being written to disk. This behaviour is common for processes such as editors, web downloads, email clients, system logging, compression programs, as well as the output from crypto-ransomware programs. These tests will use both the content of the file being written as well as metrics derived from the file's metadata such as file name and extension.

The NapierOne~\cite{Davies2021a} data set was leveraged in many of the tests that rely on file analysis. This data set is ideally suited for this task as it contains many examples of the most commonly used file types. The data set contains 5,000 example files for each of the prevalent file types shown in Table~\ref{tab:napieronetypes}. 

\begin{small}
\begin{table}
\setlength{\tabcolsep}{4pt}
\caption{NapierOne File Types}
\centering
\begin{tabular}{llllllll}
\toprule
\textbf{Type} & \textbf{} & \textbf{Type} & \textbf{} & \textbf{Type}& \textbf{} & \textbf{Type} \\
\midrule
7ZIP &  & EPS  &  & MP3  & & SVG  \\
APK  &  & EPUB &  & MP3  & & RAR  \\
BIN  &  & EXE  &  & MP4  & & TIF  \\
BMP  &  & GIF  &  & ODS  & & TXT  \\
CSS  &  & GZIP &  & OXPS & & WEBP \\
CSV  &  & HTML &  & PDF  & & XLS  \\
DLL  &  & ICS  &  & PNG  & & XLSX \\
DOC  &  & JS   &  & PS   & & XML  \\
DOCX &  & JPG  &  & PPT  & & ZIP  \\
DWG  &  & JSON &  & PPTX & &      \\
ELF  &  & MKV  &  & RAND & &     \\
\bottomrule
\end{tabular}
\label{tab:napieronetypes}
\end{table}
\end{small}

\begin{small}
\begin{table}[hbp!]
\setlength{\tabcolsep}{1.8pt}
\caption{NapierOne Ransomware Strains}
\centering
\begin{tabular}{llll}
\toprule
\textbf{Strain} & \textbf{Strain} & \textbf{Strain} \\
\midrule
AVOSLOCKER   &  DARKSIDE       & PHOBOS      \\
BADRABBIT    &   DHARMA        & RAGNAR      \\
BLACKBASTA   &   GANDCRAB      & RANSOMEX    \\
BLACKCAT     &   HELLOKITTY    & RYUK        \\
BLACKMATTER  &   JIGSAW        & SODINOKIBI  \\
CERBER       &   LOCKBIT       & SUNCRYPT    \\
CHIMERA      &   LORENZ        & TESLACRYPT  \\
CLOP         &   MAZE          & WANNACRY    \\
CONTI        &   MEDUSALOCKER  & WASTEDLOCKER\\
CRYPTOLOCKER &   NETWALKER     &             \\
CUBA         &   NOTPETYA      &             \\
\bottomrule
\end{tabular}
\label{tab:ransomwarestrains}
\end{table}
\end{small}

Apart from the normal file types found in typical use, the NapierOne data set also contains example files that have been encrypted by the ransomware strains shown in Table~\ref{tab:ransomwarestrains}. The data set contains 5,000 example encrypted files for each of these ransomware strains. (The SHA256 hash values for these ransomware strains are provided in Table~\ref{tab:ransomwarehash} which appears in the Appendix). According to previous work~\cite{Scaife2016,Nieuwenhuizen} the use of diverse families of ransomware strains is more important than the number of ransomware samples from a few families for evaluating the performance of ransomware detectors. It is because the core behavioural traits shown by crypto-ransomware in encrypting data attack does not change from one variant to the other within a family~\cite{Scaife2016}.

The entire dataset used during this research contains 365,000 files covering 73 separate and distinct file types and is publicly accessible at www.napierone.com. The dataset contains 210,000 benign files from the 42 different file types shown in Table~\ref{tab:napieronetypes} and 155,000 encrypted files from the 31 ransomware strains shown in Table~\ref{tab:ransomwarestrains}.\\

\noindent \textbf{File Magic Number Test.} Magic numbers are usually the first few bytes of a file. These are normally unique to a file format and can be used to identify many common types of files~\cite{wikimagicnumber}. While not all files contain this signature, for example, plain text files such as CSS, CSV, JSON, SVG, TXT and XLST, file types such as DOCX, PDF, XLSX and many others do contain this unique value. An extensive search was performed in an attempt to generate a comprehensive list of commonly used file types~\cite{Davies2022,wikimagicnumber,buchanan,Google2015,garrykessler,leommoore,wikilistoffilesignatures} and where possible the corresponding magic number and typical file extension for that type. This research resulted in the creation of a reference list of more than 600 entries of documented magic numbers and corresponding file extensions.

This test focuses on determining the magic number of the file under investigation and then comparing it with the file name's extension to confirm that they correlate. As plain text files do not have a magic number, then these were excluded from this test. The test was then applied to all the remaining files within the test dataset. For a file under test, if its magic number matched the corresponding expected file extension, the test passed and the file was considered benign, otherwise, the test failed and the file was considered a possible consequence of  malicious activity.\\

\noindent \textbf{Printable Characters Test.} This is a complimentary test and is only run on files that do not usually contain a magic number. As these are plain text files, then the majority of their contents should contain printable ASCII characters. Examples of files of this type are markup files such as HTML or plain text documents such as TXT. The definition of printable characters are characters that have an ASCII value between 32 and 126 as well as the format control characters which have ASCII values between nine and 13. From analysing the nearly 50,000 plain text files in the NapierOne dataset, it was found that on average plain text files contain at least 98\% printable ASCII content.

The test was then applied to all the plain text files within the test dataset. For a file under test, if its printable ASCII content was above 98\%, then the test passed and the file was considered benign, otherwise, the test failed and the file was considered a product of malicious activity.\\

\noindent \textbf{File Entropy Test.} A reoccurring theme within many crypto-ransomware detection techniques is the concept of randomness and file entropy. Researchers assert that a good indicator ~\cite{mcintosh,Genc2018,Genc2019,Kharraz2017} of crypto-ransomware activity is the generation of files whose contents appears to be random and contain no distinguishable structure. It is agreed that Well-encrypted data should be indistinguishable from random data~\cite{Choudhury2019}. Traditionally researchers in crypto-ransomware detection have chosen to use the value known as Shannon entropy~\cite{Shannon1948} when calculating this metric, however, in this research, it was decided to use the chi-square~\cite{F.R.S.2009} method of calculating this metric based on the findings of Davies~\cite{davies-entropy-2022}.

The test was then applied to all the files within the test dataset. For a file under test, if its Chi-Square entropy probability value was less than 0.01~\cite{walker2008}, then the test passed and the file was considered benign. Otherwise, the test failed and the file was considered the product of malicious activity.\\

\noindent \textbf{BitByte Value Test.} This test is based on the method described by Davies~\cite{Davies2021} which successfully distinguished between encrypted files and all other file types. This method is particularly effective at differentiating between encrypted and compressed files. A separation which previously has been proven in the past to be problematic to achieve with a reasonable level of accuracy. Essentially this test is performed by profiling the entropy distribution of the first few hundred bytes of the file under examination and comparing this profile with the entropy distribution of a control file. The difference in entropy profiles is then calculated and a value known as a~\emph{BitByte} value is determined. Files that produce lower BitByte values have a higher probability that their contents are encrypted. The research~\cite{Davies2021} identified that any BitByte value below 56, indicates with high probability, that the file is encrypted and thus possibly a consequence of a ransomware infection.

The test was then applied to all the files within the test dataset. For a file under test, if its BitByte value was greater than 56, then the test passed and the file was considered benign. Otherwise, the test failed and the file was considered a product of malicious activity.\\

\noindent \textbf{Ransom Note Creation Test.} During a crypto ransomware attack, one action often performed by the malicious process is to generate a \emph{Ransom note} file. The purpose of this file generation is two-fold. Firstly, to inform the user that their files have been encrypted and that they are the victim of a ransomware attack. Secondly, the file's contents will usually provide the victim with instructions on how they can recover from the attack and retrieve their files. The \emph{Ransom note} normally explains how the victim should transfer a specific amount of crypto-currency to the perpetrator of the attack in exchange for help in recovering the affected files. There are normally several characteristics of this \emph{Ransom note} file that can be used to distinguish it from other files. The file is normally below one KB in size, is plain text and usually contains some specific keywords such as: \emph{encrypted, ransom, tor, onion, recover, wallet, bitcoin} ~\cite{Lemmou2021c}. In this test, the actual file name is also analysed for typical ransom note file name strings such as:\emph{decrypt, readme, restore and helpme}. It has been identified that often these ransom note files are created prior to the actual encryption of the target files, so the identification of the creation of ransom notes would thus prove to be a good predictor of impending file encryption. This approach was leveraged in the HelDroid~\cite{andronio2015} ransomware detection system and utilised a text classifier that applies linguistic features to detect threatening text.

The test was then applied to all the files within the test dataset. For a file under test, if it is of limited size and its contents contain one or more of the trigger keywords, then the test failed and the file is considered malicious. Otherwise, the test passed and the file was considered benign.

\subsection{File Name Analysis}
\noindent This collection of tests is performed on the actual string value of the name of the file being written. It has been a well-known phenomenon from crypto-ransomware attacks that as well as encrypting the file contents, in the majority of cases, the affected file names will also be modified. For example by adding an extra extension or changing the original file's name. This set of tests focuses on attempting to identify this change and will again leverage the content of the NapierOne data set.\\

\noindent \textbf{File Name Entropy Test.} This test calculates the Shannon~\cite{Shannon1948} entropy value of the entire file name including any extensions that it may have. In normal operation, users tend to use lower entropy strings when naming their files.
An analysis of the original file names used to populate the NapierOne dataset shows that the average Shannon entropy of a file name is below six bits. This calculated value proves to be also language-independent~\cite{Li2005}. In many cases, when ransomware alters the contents of a file, it also alters the name of the file. Common ransomware file name manipulations are the addition of random strings to the name or its extension. This action would then increase the entropy of the affected file's name.

The test was then applied to all the files within the test dataset, using their original file names. With regards to the files generated from the execution of ransomware, then the filename generated by the ransomware was used. For a file under test, if the calculated entropy value of the entire filename string is under six bits then the test passed and the file was considered benign, otherwise, the test failed and the file was considered malicious.\\

\noindent \textbf{Known File Name Extension Test.} As mentioned above, when ransomware encrypts a file it often also tends to change or append an extra extension to the affected file. Sometimes the text of this new extension relates to the name of the ransomware but often the extension is a random string of between three and 50 characters in length. In normal operation, it is very rare that a file's extension is not a well-known value, as typically well-known applications generate files with well-known extensions. This test is aimed at checking and confirming that the extension of the file being written is one of the common extensions~\cite{Davies2022,buchanan,Google2015,leommoore}. This test uses the collated list, created by the authors, of known extensions which are also used in the \emph{Magic Number Test} described in Section~\ref{filecontentanalysis}. If the file extension is present in the list, then it is considered to be well-known. If the file name contains multiple extensions, then this test is applied to the last extension. 

The test was then applied to all the files within the test dataset. For a file under test, if the file's extension is well-known then the test passed and the file was considered benign, otherwise, the test failed and the file was considered malicious.\\

\noindent \textbf{File Name Extension Entropy Test.} This test calculates the Shannon~\cite{Shannon1948} entropy value of the file name's extension. If the file has multiple extensions, then the entropy of the entire extension chain is calculated. Often crypto-ransomware will append an extra extension to a file that it has encrypted. This extension can be a text string relating to the ransomware strain, but more recently it has been a random string of between three and 50 characters. When analysing the entropy value of all the extensions in the list of well-known extensions it was found that they all had a Shannon entropy value of below six bits. 

The test was then applied to all the files within the test dataset. For a file under test, if the calculated entropy value of the file's extension, or extensions, is below six bytes then the test passed and the file was considered benign, otherwise, the test failed and the file was considered malicious.\\

\subsection{Executable Analysis}
\noindent The collection of tests described in this section relates to tests performed on the executable code files used to launch the process as well as tests performed on a process's memory captured during its execution. Benign programs were selected that would normally generate files of a specific type. Specific details of the benign programs analysed are provided in Table~\ref{tab:benignprograms}. For example, files of type DOCX would usually be created using the Microsoft Word application, so the executable for this application was analysed as well as its memory during its execution.\\

\noindent \textbf{Strings in Executable Test.}
\noindent Often ransomware executables contain anti-analysis techniques in an attempt to prevent researchers from inspecting the content of the code. These techniques can include obfuscation, polymorphism and encryption of the content of the executable. A consequence of this is that the number of humanly readable strings found within such a file could be significantly lower than would normally be expected. This static analysis technique was applied to both benign as well as ransomware executable files and took the form of extracting strings from the executable and then counting the number and frequency of Windows Application Programming Interface(API) strings that could be identified. This technique has also been leveraged in other ransomware detection systems such as R-PackDroid~\cite{packdroid}.

No specific metrics, such as the expected number of API strings per KB, are currently available in the literature. So these tests are more exploratory to discover if the type and frequency of API calls differ significantly between ransomware and benign executables and if this measurement would be a useful contributor to a malice score calculation in a ransomware detection system.\\

\noindent \textbf{Creation and Modification Dates Test.}
\noindent Executable files normally have a significant time interval between when they were placed on the file system and the current execution time. A small interval between the creation date and time and the current date and time could also be used as an indicator of a recently placed malicious program.

This static analysis test was applied to all the executable files shown in the appendix in Tables~\ref{tab:ransomwarehash}~and~\ref{tab:benignprograms}. For an executable file under test, if the file's creation or modification date is greater than one day then the test passed and the executable file was considered benign, otherwise, the test failed and the file was considered malicious.

\subsubsection{Process Analysis}
\noindent The following tests could be performed on running processes to determine if any indicators could be identified, that would suggest that the process was malicious.
The memory contents of the process under investigation are analysed for indicators of malicious behaviour.\\

\noindent \textbf{File-less Execution Test.}
Running processes that do not have an underlying executable on the file system could be considered suspicious as some forms of ransomware execute by being directly injected into memory. These injected programs would then have no underlying executable file present on the file system. This is unusual behaviour for a process and can be used to flag irregular behaviour~\cite{KARA2023119133}.

This test was applied to the running process. If the process is associated with a file on the file system then the test passed and the process file was considered benign, otherwise, the test failed and the process was considered malicious.\\

\noindent \textbf{Cryptographic Key Identification Test.}
The memory and underlying executable file used to launch the process under investigation will be examined for traces of cryptographic keys, as these could indicate that the process is, or will shortly begin, encrypting files. The memory will be searched for keys for the following three cryptographic algorithms: AES~\cite{Balogh2011,Halderman2009a,Maartmann-Moe2009b,Heninger2008,Kornblum}, Salsa20~\cite{salsa} and RSA~\cite{Joseph2020a,Klein2006}. The AES key testing included checking for the presence of keys of length 128, 192 and 256 bits.

Initially, the executable file that will be used to launch the process with be examined. Subsequently, the memory of the running process will be checked on two occasions, firstly, directly after the process has launched and then subsequently checked again 30 seconds after launch. If no keys are found in each of these tests, then the test passed and the process was considered benign, otherwise, if keys are discovered, the test failed and the process was considered malicious.\\

\noindent \textbf{Ransom Note Identification Test.}
The memory of the process under investigation will be examined for traces of typical strings that often appear within ransom notes. These are files normally generated by ransomware programs and are used to inform the user that they have been the victim of a ransomware attack. These files usually contain information on how the user may recover their data. The presence of many keywords close together within the process's memory would be an indicator that the process could be malicious. This test is similar to the previous Ransom Note Creation Test, using the same keywords, however, in this case, it will be performed on the process's memory and not on its output.

This test was applied to the running process. If the process's memory does not contain several examples of the keywords, then the test passed and the process was considered benign, otherwise, the test failed and the process was considered malicious.\\

\noindent \textbf{Windows API Analysis Test.}
The memory of the process under investigation will be examined and a review of the number and frequency of all the found window's application programming interface (API) calls will be performed. Executables use these API calls to interact with the operating system and the number and type of calls used together with their frequency will be investigated to determine if this could be used as an indicator that the process under investigation is malicious. This test is similar to the previous Strings in Executable Test, however, in this case, it will be performed on the process's memory and not on the executable file used to launch the process.

No specific metrics, such as the expected number of API strings per KB, are currently available in the literature. So these tests are more exploratory to discover if the type and frequency of API calls differ significantly between ransomware and benign executables and if this measurement would be a useful contributor to a malice score calculation in a ransomware detection system.

\subsection{Behaviour Analysis}
\noindent The actions and behaviour exhibited by the ransomware can also be monitored to identify suspicious behaviour. These tests are outlined below.\\

\noindent \textbf{Modification of System Restore Points.}
System restore points are used to recover a system's state or file system files. There are very few occasions where a process needs to issue commands relating to system restore points, especially concerning their deletion. The state of the system's restore points will be monitored, during the execution of the process under investigation, to determine if they are modified.

This test was applied to the running process. If the systems restore points remained intact two minutes after the launch of the process, then the test passed and the process was considered benign, otherwise, if the restore points had been altered or deleted, the test failed and the process was considered malicious.\\

\noindent \textbf{Process escalation}
Some ransomware processes attempt to gain elevated access to resources that are normally protected from an application or user. This is attempted so that the process can gain deeper and broader control of the system and allow them to perform more destructive actions. Identification of such behaviour would prove to be a useful indicator of malicious activity.

This test was applied to the running process. If the running process achieves elevated access or spawns a child process with elevated access then the test fails and the process is considered malicious, otherwise, if the access remains unchanged then the test passed and the process was considered benign.

\section{Evaluation and Discussion}
\label{cha:evaluation}
\begin{figure*}
  \includegraphics[scale=0.665]{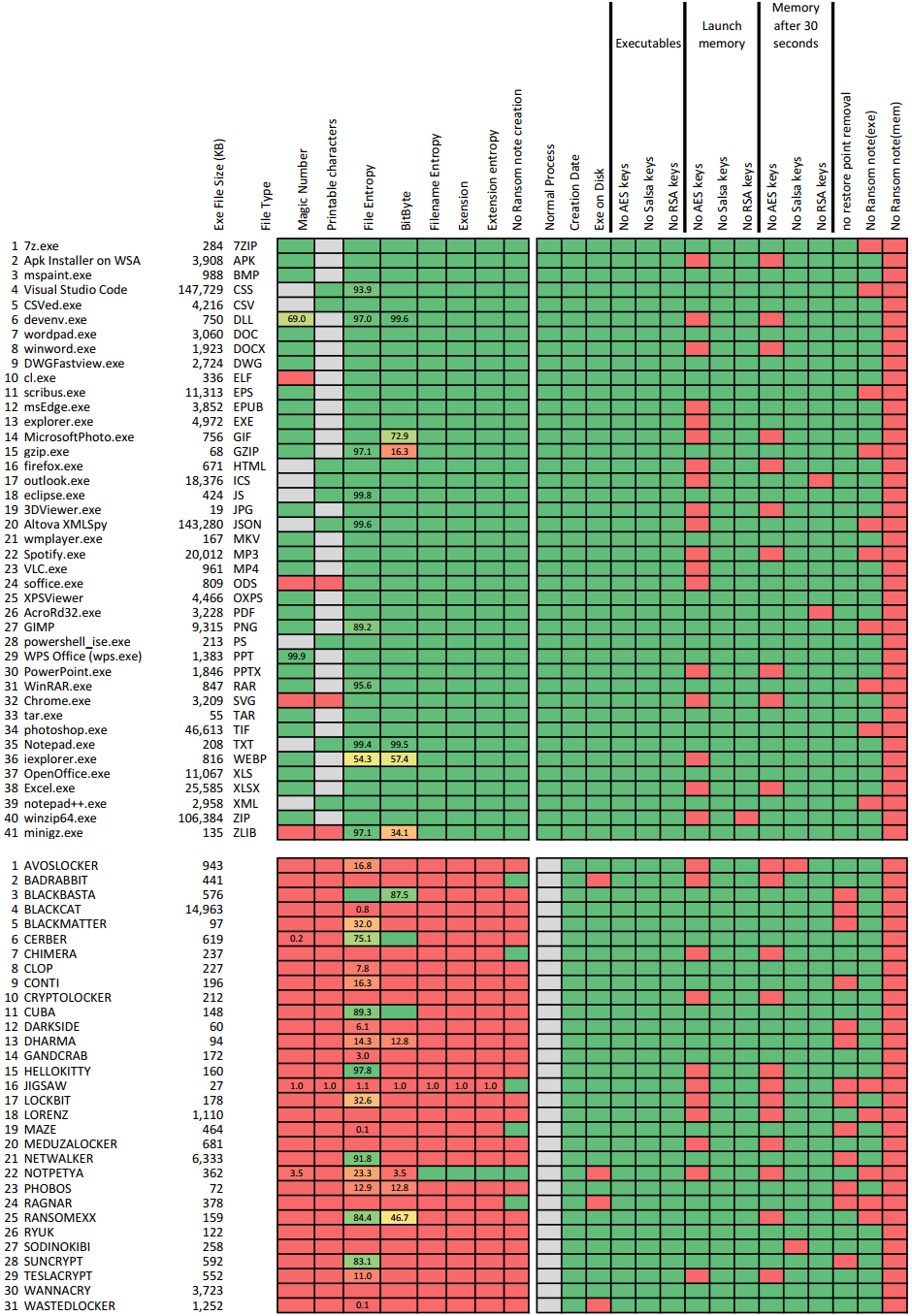}
  \caption{Results Overview.}
  \label{fig:results-grid}
\end{figure*}

\noindent The majority of the recorded results for the tests described in Section~\ref{cha:methodology} are provided in Figure~\ref{fig:results-grid}.
The cell colours represent the success of the test and are graded from green to red. 100\% pass rate results are represented as a dark green colour, the colour changes depending on the success rate to red which indicates 0\% pass rate, or alternatively 100\% failure rate. Where the colour does not clearly show the result, then the percentage number is also displayed. Grey indicates that the specific test was not executed on that particular file type. For example, as mentioned above, if the file type should contain a magic number, then this test was performed and the printable character test was ignored.
\begin{table}
\caption{Possible classification outcomes}
\centering
\begin{tabular}{l@{}|l}
\hline
 \textbf{Classification} & \textbf{Description}\\
\hline
True Positive\hspace{3.3mm}(TP) & Test passes on benign file\\
True Negative\hspace{2.0mm}(TN) & Test fails on ransomware file \\
False Positive\hspace{2.5mm}(FP) & Test passes on ransomware file \\
False Negative\hspace{1mm}(FN) & Test fails on benign file \\
\hline
\end{tabular}
\label{tab:classification-options}
\end{table}

\begin{figure}
  \includegraphics[scale=0.50]{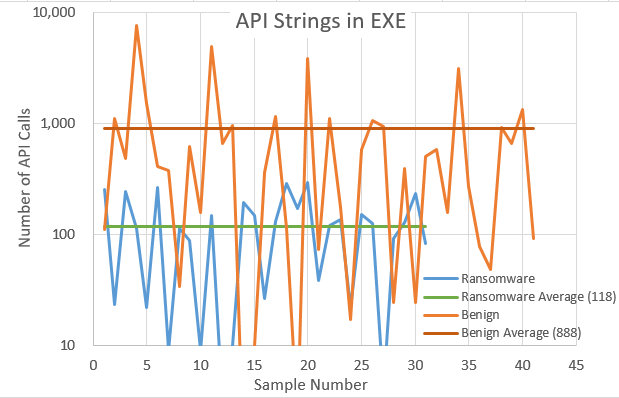}
  \caption{Total API Calls in executable}
  \label{fig:api-exe}
\end{figure}

\begin{figure}
  \includegraphics[scale=0.50]{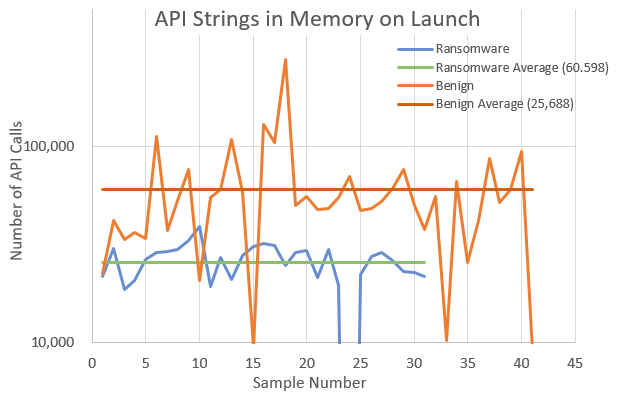}
  \caption{Total API Calls in memory}
  \label{fig:api-mem}
\end{figure}

\begin{figure}
  \includegraphics[scale=0.5]{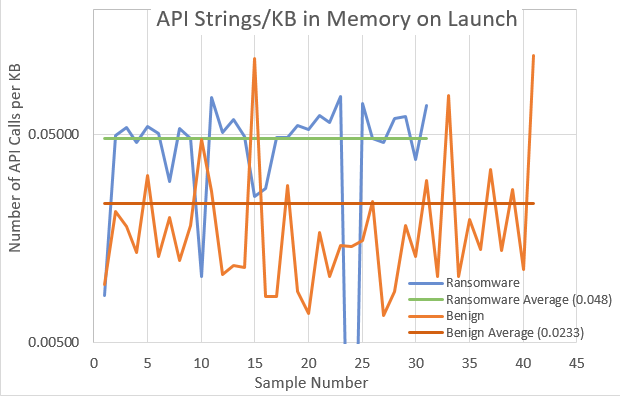}
  \caption{Average API Calls in Memory per KB at Launch}
  \label{fig:avg-api-mem-launch}
\end{figure}

\begin{figure}
  \includegraphics[scale=0.47]{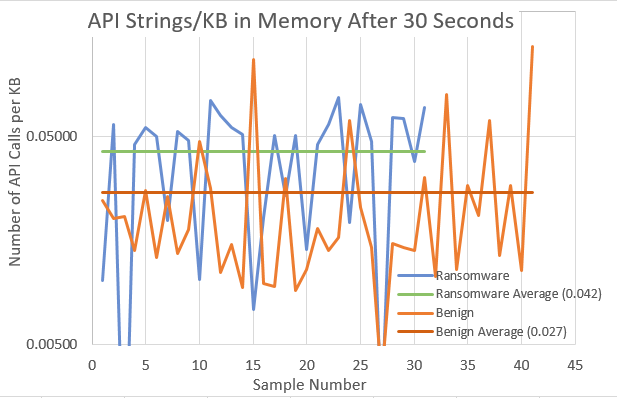}
  \caption{Average API Calls in Memory per KB At 30 Seconds}
  \label{fig:avg-api-mem-30s}
\end{figure}

Some of the tests were exploratory in nature in an attempt to discover if the gathered metrics could be used to identify malicious code. Examples of these exploratory tests were in the cataloguing of API calls distinguishable in the executable as well as the process memory directly after launch and then again 30 seconds after launch. The results of these tests are presented in Figures~\ref{fig:api-exe}-\ref{fig:avg-api-mem-30s}. The remainder of this section reviews the results gathered during the testing and provides some context, discussion and background into the tests and the recorded results. A clarification of a test's success is provided in Table~\ref{tab:classification-options}.\\

\begin{table}
\centering
\caption{File Test Performance Metrics}
\begin{tabular}{lcccc}
\hline
 & Accuracy & Recall & Precision & F1 \\
 \hline
Magic number    & 0.961 & 0.998 & 0.923 & 0.959 \\
Printable Char. & 0.999 & 0.999 & 0.999 & 0.999 \\
File Entropy    & 0.865 & 0.831 & 0.958 & 0.890 \\
BitByte         & 0.919 & 0.914 & 0.946 & 0.930 \\
Filename Ent.   & 0.999 & 0.999 & 0.999 & 0.999 \\
Extension       & 0.999 & 0.999 & 0.999 & 0.999 \\
Extension Ent.  & 0.999 & 0.999 & 0.999 & 0.999 \\
\hline
\end{tabular}
\label{tab:enhancedconfusion-tab}
\end{table}

\noindent \textbf{File content analysis.}
These tests were performed on the files generated by a process. These tests included the analysis of the created file's magic number value, or for plain text files, the percentage of humanly readable characters within the file was analysed. Other tests included the Chi-Square entropy of the content of the file as well as the BitByte value test. Of all the tests performed these were some of the most successful in differentiating between output generated from benign and malicious processes, a summary of the results is provided in Table~\ref{tab:enhancedconfusion-tab}. No individual test achieved 100\% accuracy but the BitByte test is worth highlighting as its results were more accurate than the plain entropy tests when working on files with unknown content. The magic number test combined with known file extension tests also achieved high accuracy, but these rely on the created files having a known extension. These tests could be bypassed, by ransomware using well-known extensions on their output, as highlighted by the results recorded when analysing the files generated by the NotPetya ransomware strain which does not modify the extension of the files it attacks~\cite{Sai2019}.
\\

\noindent \textbf{Generated file name analysis.}
These tests were performed on the names of the files generated by the process. These tests included the analysis of the entropy of the entire filename, the entropy of the file name's extension as well as validating if the file name's extension was a known value. These tests also achieved high accuracy, a summary of the results is provided in Table~\ref{tab:enhancedconfusion-tab}. Contributing factors to this high accuracy was that the benign files used all had well-known file extensions and almost all the tested ransomware strains modified the files name and/or extension in a way that increased the overall entropy of the filename. While the testing did cover more than 30 different ransomware strains, it may not be sufficiently broad enough to generalise this phenomenon. As with some of the file content tests, the exception being the files generated by the NotPetya ransomware strain, which was able to successfully evade this group of tests. This leads us to think that these tests should be applied to a larger test dataset, before generalising the findings.
\\

\noindent \textbf{Ransomnote tests.}
These tests can also be divided into static and dynamic analysis tests. The static portion of the tests involved examining the executable file used to launch the process and trying to identify several occurrences of typical strings used in ransom notes. This analysis could be performed prior to the launching of the process. In one of the dynamic analysis tests, the running process's volatile memory was examined for the existence of these same ransom note strings. In the other dynamic analysis test, the files generated by the process were examined to determine if the file being created could possibly be a ransom note. No ransom note strings were found in either the begin or ransomware executable binaries. The success rate when looking for ransom note strings within the memory was very low with positive matches only 20\% of the time. These matches were relatively evenly distributed between benign and malicious programs. A consequence of this is that it seems that these metrics would not be suitable for use within a ransomware detection program. The accuracy may be improved for these tests by possibly applying some additional logic to the search, for example, by increasing the dictionary of keywords being searched for, applying natural language processing on the found strings, or analysing the distance between where these words appear and applying a ranking or weighting to the found strings.

The results regarding the dynamic test of analysing the contents of files being created by the process were much more encouraging. No files generated by the benign programs were marked as ransomware, and 80\% of the ransom notes generated by the ransomware were successfully identified. Some reasons why this rate was not even higher were that some ransomware strains do not create ransom notes, some ransom notes were actual graphics and some ransomware strains changed the desktop background to display the ransom message. This is a promising finding as many ransomware strains create the ransom note prior to the encryption~\cite{Lemmou2021c} of the data and a successful interception at this point in the attack would be beneficial.
\\

\noindent \textbf{Identification of cryptographic artefacts.}
These tests involved attempting to identify cryptographic algorithm artefacts using both static and dynamic analysis methods. The static portion of the tests involved looking for these artefacts in the executable binary files used to launch the process. This analysis could be performed prior to the launching of the process. The dynamic aspect of these tests involved looking for these artefacts in the process's volatile memory, precisely after the process has been launched and then again 30 seconds after the process's launch. The artefacts that were being searched for were AES encryption algorithm keys of length 128 bits, 192 bits and 256 bits, as well as RSA asymmetric and Salsa20 encryption algorithm keys. This resulted in 15 distinct tests per file type resulting in a total of more than 1,000 tests being conducted for this group of tests. When reviewing the results it can be seen that no cryptographic artefacts were identifiable in any of the test binaries. Regarding the AES key discovery, then no real pattern could be identified. For benign programs, these keys were identified in 44\% of the samples and with ransomware programs, these keys were identified in 35\% of the samples. These findings indicate that this metric in its current format is not particularly suited for indicating ransomware activity. It is a known behaviour, that ransomware does use cryptography during its execution, so some explanation for the lack of successful key identification could be that these algorithms are not used in these ransomware samples or that the encryption has either not commenced or has completed when the analysis was performed. The presence of these artefacts in memory of benign programs is not ideal as it complicates the metric.\\

\noindent \textbf{Behavioural analysis.} 
These tests are aimed at \newline analysing the behaviour of the process under investigation. The idea behind the first test, normal process, was to try and identify if a running process attempts to alter its execution privileges and try and run as an elevated user. None of the benign processes did this, however, many of the ransomware samples used, would not execute correctly without them being started as the administrator user, which negated the usefulness of this test. No identifying trend could be used to differentiate benign and malicious binary files using the file creation date. With regards to file-less execution, four ransomware samples did spawn a malicious process that had no underlying binary file on disk and approximately 45\% of ransomware programs removed system restore points soon after they started executing. Both these last two behaviours were only observed with ransomware programs and could be used as a contributing factor when trying to determine if the process is malicious or not.
\\

\noindent \textbf{Analysis Tests.} 
Some exploratory investigation was also performed in cataloguing and analysing the number and frequency of standard Windows API calls within both the binary executable file (static analysis) as well as the process's volatile memory (dynamic analysis) directly after launch and then again 30 seconds after launch. Note that the Y-axis on the following figures has a logarithmic scale.
From Figure~\ref{fig:api-exe}, it can be seen that the number of API calls present in the executables of benign programs differ by a factor of eight when compared to the number of API calls identified within ransomware programs  One possible explanation for this could be that ransomware programs often try and obfuscate their structure prior to execution in an attempt to hinder analysis and a consequence of this being that the API calls are hidden.

To normalise these results, the values were then plotted as a ratio of the number of API calls present divided by the analysed executable file size. These normalised results are shown in Figure~\ref{fig:api-mem}
The programs were then launched and the volatile memory used by each of these programs was then captured and analysed for Windows API calls. A comparison of the API calls present within each process's memory is presented in Figure~\ref{fig:avg-api-mem-launch}. Again to aid comparison, the graphs have been normalised by dividing the total number of calls by the size of the total memory being used.
Finally, the launched program's memory was captured again 30 seconds after launch and analysed for Windows API calls. A comparison of the API calls present within each process's memory is presented in Figure~\ref{fig:avg-api-mem-30s} with the results being normalised.

When reviewing the captured results, it can be seen that the identified API calls within the binary files show signs of possibly being a useful indicator of Windows API obfuscation and thus an indicator of a possibly malicious program. The measurements show that there is an obvious difference between the number of API calls found within the benign and malicious executables. The difference between these two types of executables is not so prominent when analysing the process's volatile memory. However, it is felt by the researchers that these findings merit further investigation. As they stand, using the current metrics, these results would not prove useful as a contributor to the suite of tests used in the malice score calculation. Some refinement of the measurement, such as targeting specific API calls or call frequency analysis may enhance the accuracy of this type of measurement and further investigation into this would be beneficial.

\subsection{Majority Voting}
\noindent When reviewing the results from the separate tests mentioned above, it can be seen that several tests achieved a high degree of accuracy in differentiating between benign and malicious programs, using both static and dynamic tests. The results from some tests such as attempting to identify cryptographic artefacts, ransom note identification within the process and executable or Windows API enumeration, delivered inconclusive results and these tests would require some more investigation, analysis and modification.

\begin{figure}
  \includegraphics[scale=0.34]{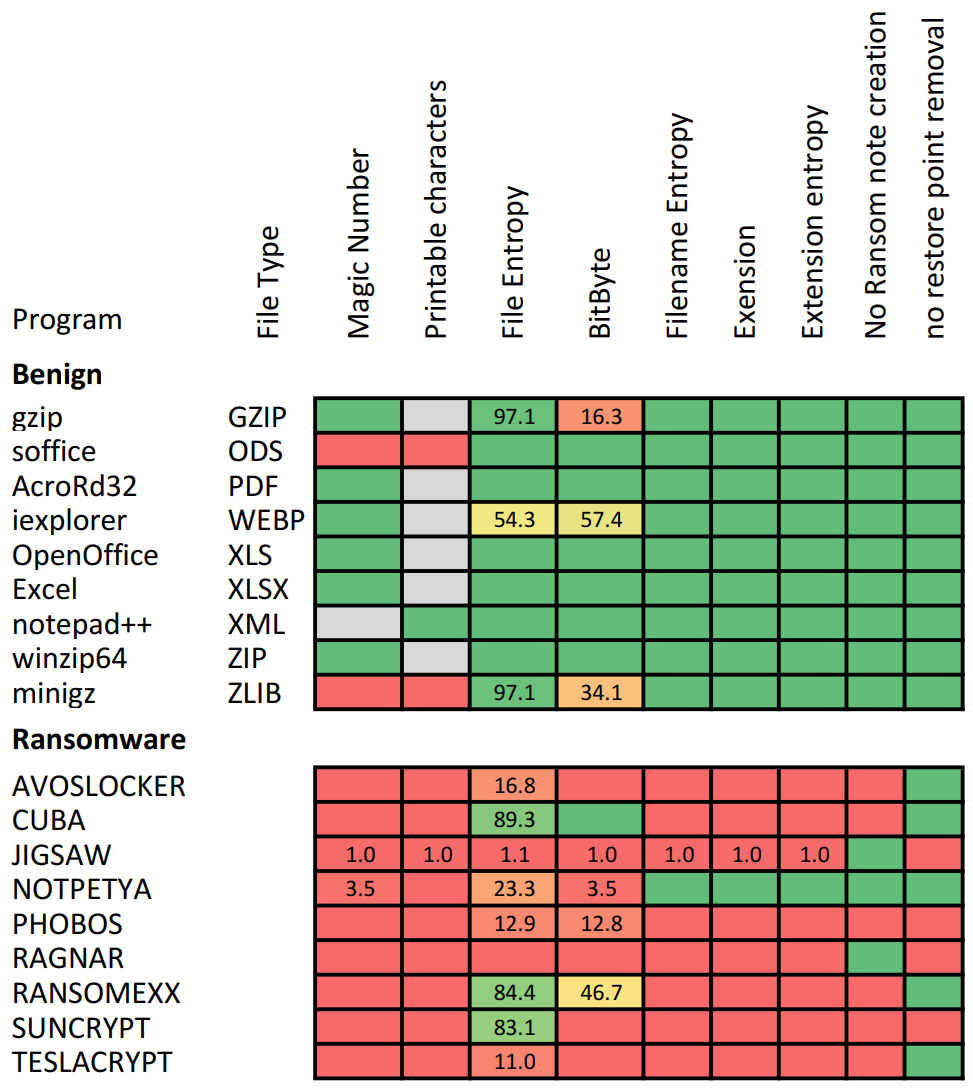}
  \caption{Interesting Results.}
  \label{fig:results-specific}
\end{figure}

It is proposed that a system could be developed that uses a combination of the tests that have been found to be accurate in identifying ransomware. Each test's vote would contribute to an overall malice score for the target file or process, and based on the maximum number of votes the system would classify the target as either malicious or benign. For example, a system could be developed that used the following tests: 
created file name and extension entropy, well-known extensions, file magic number and printable characters, file content BitByte and entropy values, ransom note creation detection and system restore point removal detection. Based on the findings from this research, a system configured with these tests would have an accuracy of 0.9989 on the dataset used. Some of the interesting test results are highlighted in Figure~\ref{fig:results-specific}.

When reviewing the results for benign programs shown in Figure~\ref{fig:results-specific}, it can be seen that the majority of tests consider the processes/files to be benign. Even in cases where some of the individual tests do occasionally give false positives, in all cases, the majority of the tests vote correctly resulting in a correct overall classification. For example, when looking at the classification for WEBP file types, it can be seen that the individual file entropy and BitByte tests, result in a classification accuracy of around 55\%, and 45\% of the samples are incorrectly classified as malicious. However, as the remaining six tests correctly vote that the file is benign, these files are ultimately classified as benign. Likewise, when reviewing the result for the ransomware files, in most cases the majority of tests classify the file/process as malicious. The only exception would be on the very rare occasion, the files generated by the Jigsaw ransomware strain, may theoretically receive a false positive classification if the majority of the tests vote that the file/process is benign. 

 A major strength of the majority voting approach to ransomware detection is that not every test needs to correctly classify a malicious program every time. With equal weighting on the result of each test, it would be sufficient for just a majority of tests to correctly classify the target, for the system to work successfully. Some work could also be performed to investigate whether a weighting or bias could also be applied to the test results meaning that some tests would then have a greater influence on the overall outcome of the classification, than others.

As the detection technique relies on well-known discrete tests, it is also easier for the detection model to be modified, updated and tuned as opposed to a machine learning model where the weightings and strengths of the learned model can be unknown or difficult to influence.
\section{Conclusion}
\label{cha:conclusion}
\noindent This paper proposes a ransomware detection system using a majority voting-based approach.
A final \emph{malice score} is derived from the combination of the results from many discrete tests that are conducted on the target process, its executable file or the output that the process generates. These distinct results are then aggregated and used as input for the malice score generation. Based on this score the target is classified as benign or malicious. The paper proposes 23 main tests that could potentially be used in a ransomware detection system with their outcomes, contributing to the overall malice score. The paper also investigates additional potential metrics that could be used in ransomware detection, for example, the presence of Windows API calls in the binary and executing processes' volatile memory.

This research demonstrates that many of the proposed tests achieved a high degree of accuracy in differentiating between benign and malicious targets. The accuracy was then enhanced when a selection of these tests was then combined into a majority voting model. One proposed majority voting model achieves an accuracy of 0.9989. The collaborative approach in generating the final result has many advantages, for example, some individual tests on some occasions may produce incorrect classifications, but the overall accuracy of the detection system as a whole will be unaffected if the majority of the tests produce the correct results.

As this majority voting detection technique relies on well-known and easily understandable discrete tests, then it is easier for the model to be modified, updated and tuned as opposed to a machine learning approach where the weightings and strengths of the learned model can be unknown or difficult to influence. An additional advantage is that while machine learning models require training, the majority voting approach, proposed in this paper, does not. 

\subsection{Limitations}
\label{limitations}
\noindent While the majority voting approach to identifying malicious processes has a high level of accuracy, as always the situation exists where once a ransomware developer is aware of the techniques being used to identify malicious behaviour, they have the possibility of modifying or adapting the ransomware's behaviour to avoid the tests in newer releases of their programs. The advantage of the majority voting approach is that the system does not rely on a single catch-all test, rather detection is a combination of many accurate tests. A consequence of this is that the ransomware developer may have to significantly modify the behaviour of their programs, and possibly disregard some aspects of their original behaviour to avoid detection.

\subsection{Future Work}
\label{future}
\noindent The results achieved during the Windows API call analysis could possibly be improved by further investigation and modifications to the types of API calls present, their frequency and their position within the file or process memory. One area of further work would be a deeper analysis of this aspect of the binaries and volatile memory.
Another area of work would be to introduce a weighting element to the measurements, allowing some tests to have a greater influence on the final classification results.

Analyses of other types of tests could also be performed. Examples of which could be: multiple-file read and write operations, high entropy differences between read and write operations, file tree traversal, privilege escalation, accessing crypto API functionality, accessing unusual domain names, generation of large amounts of traffic, DGA detection~\cite{chadha,salehi} and the termination of a large number of processes.

\printcredits

\bibliographystyle{cas-model2-names}

\bibliography{bibliography}

\begin{thebibliography}{77}
\expandafter\ifx\csname natexlab\endcsname\relax\def\natexlab#1{#1}\fi
\providecommand{\url}[1]{\texttt{#1}}
\providecommand{\urldate}[1]{\texttt{#1}}
\providecommand{\href}[2]{#2}
\providecommand{\path}[1]{#1}
\providecommand{\DOIprefix}{doi:}
\providecommand{\ArXivprefix}{arXiv:}
\providecommand{\URLprefix}{URL: }
\providecommand{\URLDATEprefix}{(Last Accessed: }
\providecommand{\Pubmedprefix}{pmid:}
\providecommand{\doi}[1]{\href{http://dx.doi.org/#1}{\path{#1}}}
\providecommand{\Pubmed}[1]{\href{pmid:#1}{\path{#1}}}
\providecommand{\bibinfo}[2]{#2}
\ifx\xfnm\relax \def\xfnm[#1]{\unskip,\space#1}\fi
\bibitem[{Abbasi et~al.(2021)Abbasi, Al-Sahaf and Welch}]{abbasi}
\bibinfo{author}{Abbasi, M.S.}, \bibinfo{author}{Al-Sahaf, H.},
  \bibinfo{author}{Welch, I.}, \bibinfo{year}{2021}.
\newblock \bibinfo{title}{{Automated Behavior-based Malice Scoring of
  Ransomware Using Genetic Programming}}.
\newblock \bibinfo{journal}{2021 IEEE Symposium Series on Computational
  Intelligence, SSCI 2021 - Proceedings}
  \DOIprefix\doi{10.1109/SSCI50451.2021.9660009}.
\bibitem[{Ahmed et~al.(2021)Ahmed, Kim, Camtepe and Nepal}]{Ahmed2021}
\bibinfo{author}{Ahmed, M.E.}, \bibinfo{author}{Kim, H.},
  \bibinfo{author}{Camtepe, S.}, \bibinfo{author}{Nepal, S.},
  \bibinfo{year}{2021}.
\newblock \bibinfo{title}{{Peeler: Profiling Kernel-Level Events to Detect
  Ransomware}}.
\newblock \bibinfo{journal}{Lecture Notes in Computer Science (including
  subseries Lecture Notes in Artificial Intelligence and Lecture Notes in
  Bioinformatics)} \bibinfo{volume}{12972 LNCS}, \bibinfo{pages}{240--260}.
\newblock \DOIprefix\doi{10.1007/978-3-030-88418-5\_12},
  \href{http://arxiv.org/abs/2101.12434}{\tt arXiv:2101.12434}.
\bibitem[{Ahmed et~al.(2020)Ahmed, Ko{\c{c}}er and Al-Rimy}]{Ahmed2020a}
\bibinfo{author}{Ahmed, Y.A.}, \bibinfo{author}{Ko{\c{c}}er, B.},
  \bibinfo{author}{Al-Rimy, B.A.S.}, \bibinfo{year}{2020}.
\newblock \bibinfo{title}{{Automated Analysis Approach for the Detection of
  High Survivable Ransomware}}.
\newblock \bibinfo{journal}{KSII Transactions on Internet and Information
  Systems} \bibinfo{volume}{14}, \bibinfo{pages}{2236--2257}.
\newblock \DOIprefix\doi{10.3837/tiis.2020.05.021}.
\bibitem[{Al-rimy et~al.(2019)Al-rimy, Maarof and Shaid}]{Al-rimy2019}
\bibinfo{author}{Al-rimy, B.A.S.}, \bibinfo{author}{Maarof, M.A.},
  \bibinfo{author}{Shaid, S.Z.M.}, \bibinfo{year}{2019}.
\newblock \bibinfo{title}{{Crypto-ransomware early detection model using novel
  incremental bagging with enhanced semi-random subspace selection}}.
\newblock \bibinfo{journal}{Future Generation Computer Systems}
  \bibinfo{volume}{101}, \bibinfo{pages}{476--491}.
\newblock \URLprefix \url{https://doi.org/10.1016/j.future.2019.06.005},
  \DOIprefix\doi{10.1016/j.future.2019.06.005}.
\bibitem[{Alam et~al.(2020)Alam, Sinha, Bhattacharya, Dutta, Mukhopadhyay and
  Chattopadhyay}]{Alam2020}
\bibinfo{author}{Alam, M.}, \bibinfo{author}{Sinha, S.},
  \bibinfo{author}{Bhattacharya, S.}, \bibinfo{author}{Dutta, S.},
  \bibinfo{author}{Mukhopadhyay, D.}, \bibinfo{author}{Chattopadhyay, A.},
  \bibinfo{year}{2020}.
\newblock \bibinfo{title}{{RAPPER: Ransomware Prevention via Performance
  Counters}}.
\newblock \bibinfo{journal}{Arxiv} , \bibinfo{pages}{1--18}\URLprefix
  \url{http://arxiv.org/abs/2004.01712},
  \href{http://arxiv.org/abs/2004.01712}{\tt arXiv:2004.01712}.
\bibitem[{Andronio et~al.(2015)Andronio, Zanero and Maggi}]{andronio2015}
\bibinfo{author}{Andronio, N.}, \bibinfo{author}{Zanero, S.},
  \bibinfo{author}{Maggi, F.}, \bibinfo{year}{2015}.
\newblock \bibinfo{title}{{HelDroid: Dissecting and Detecting Mobile
  Ransomware}}, in: \bibinfo{editor}{Bos, H.}, \bibinfo{editor}{Monrose, F.},
  \bibinfo{editor}{Blanc, G.} (Eds.), \bibinfo{booktitle}{Research in Attacks,
  Intrusions, and Defenses}, \bibinfo{publisher}{Springer International
  Publishing}, \bibinfo{address}{Cham}. pp. \bibinfo{pages}{382--404}.
\bibitem[{Balogh and Pondelik(2011)}]{Balogh2011}
\bibinfo{author}{Balogh, {\v{S}}.}, \bibinfo{author}{Pondelik, M.},
  \bibinfo{year}{2011}.
\newblock \bibinfo{title}{{Capturing encryption keys for digital analysis}}.
\newblock \bibinfo{journal}{Proceedings of the 6th IEEE International
  Conference on Intelligent Data Acquisition and Advanced Computing Systems:
  Technology and Applications, IDAACS'2011} \bibinfo{volume}{2},
  \bibinfo{pages}{759--763}.
\newblock \DOIprefix\doi{10.1109/IDAACS.2011.6072872}.
\bibitem[{Berrueta et~al.(2019)Berrueta, Morato, Magana and
  Izal}]{Berrueta2019}
\bibinfo{author}{Berrueta, E.}, \bibinfo{author}{Morato, D.},
  \bibinfo{author}{Magana, E.}, \bibinfo{author}{Izal, M.},
  \bibinfo{year}{2019}.
\newblock \bibinfo{title}{{A Survey on Detection Techniques for Cryptographic
  Ransomware}}.
\newblock \bibinfo{journal}{IEEE Access} \bibinfo{volume}{7},
  \bibinfo{pages}{144925--144944}.
\newblock \DOIprefix\doi{10.1109/ACCESS.2019.2945839}.
\bibitem[{Bottazzi et~al.(2018)Bottazzi, Italiano, Spera, Bottazzi, Italiano,
  Spera, Ransomware, Through, Bottazzi, Italiano and Spera}]{Bottazzi2018}
\bibinfo{author}{Bottazzi, G.}, \bibinfo{author}{Italiano, G.},
  \bibinfo{author}{Spera, D.}, \bibinfo{author}{Bottazzi, G.},
  \bibinfo{author}{Italiano, G.}, \bibinfo{author}{Spera, D.},
  \bibinfo{author}{Ransomware, P.}, \bibinfo{author}{Through, A.},
  \bibinfo{author}{Bottazzi, G.}, \bibinfo{author}{Italiano, G.F.},
  \bibinfo{author}{Spera, D.}, \bibinfo{year}{2018}.
\newblock \bibinfo{title}{{Preventing Ransomware Attacks Through File System
  Filter Drivers}}, in: \bibinfo{booktitle}{Proceedings of the Second Italian
  Conference on Cyber Security (ITASEC18)}, p.~\bibinfo{pages}{4}.
\newblock \URLprefix \url{https://www.researchgate.net/publication/323125541
  \_Preventing\_Ransomware\_Attacks\_Through\_File\_System \_Filter\_Drivers}.
\bibitem[{Buchanan()}]{buchanan}
\bibinfo{author}{Buchanan}, .
\newblock \bibinfo{title}{{Digital Forensics Magic Numbers}}.
\newblock \URLprefix \url{https://asecuritysite.com/forensics/magic},
  \URLDATEprefix \urldate{2022-10-20}).
\bibitem[{Chadha and Kumar(2017)}]{chadha}
\bibinfo{author}{Chadha, S.}, \bibinfo{author}{Kumar, U.},
  \bibinfo{year}{2017}.
\newblock \bibinfo{title}{{Ransomware: Let's fight back!}}, in:
  \bibinfo{booktitle}{2017 International Conference on Computing, Communication
  and Automation (ICCCA)}, pp. \bibinfo{pages}{925--930}.
\newblock \DOIprefix\doi{10.1109/CCAA.2017.8229926}.
\bibitem[{Choudhury et~al.(2019)Choudhury, Kumar, Nandi and
  Athithan}]{Choudhury2019}
\bibinfo{author}{Choudhury, P.}, \bibinfo{author}{Kumar, K.R.},
  \bibinfo{author}{Nandi, S.}, \bibinfo{author}{Athithan, G.},
  \bibinfo{year}{2019}.
\newblock \bibinfo{title}{{An empirical approach towards characterization of
  encrypted and unencrypted VoIP traffic}}.
\newblock \bibinfo{journal}{Multimedia Tools and Applications}
  \bibinfo{volume}{79}, \bibinfo{pages}{603--631}.
\newblock \DOIprefix\doi{10.1007/s11042-019-08088-w}.
\bibitem[{Continella et~al.(2016)Continella, Guagnelli, Zingaro, {De Pasquale},
  Barenghi, Zanero and Maggi}]{Continella}
\bibinfo{author}{Continella, A.}, \bibinfo{author}{Guagnelli, A.},
  \bibinfo{author}{Zingaro, G.}, \bibinfo{author}{{De Pasquale}, G.},
  \bibinfo{author}{Barenghi, A.}, \bibinfo{author}{Zanero, S.},
  \bibinfo{author}{Maggi, F.}, \bibinfo{year}{2016}.
\newblock \bibinfo{title}{{ShieldFS: A self-healing, ransomware-aware file
  system}}, in: \bibinfo{booktitle}{Proceedings of the 32nd Annual Conference
  on Computer Security Applications}, pp. \bibinfo{pages}{336--347}.
\newblock \URLprefix \url{https://dl.acm.org/doi/10.1145/2991079.2991110},
  \DOIprefix\doi{10.1145/2991079.2991110}.
\bibitem[{Dargahi et~al.(2019)Dargahi, Dehghantanha, Bahrami, Conti, Bianchi
  and Benedetto}]{Dargahi2019}
\bibinfo{author}{Dargahi, T.}, \bibinfo{author}{Dehghantanha, A.},
  \bibinfo{author}{Bahrami, P.N.}, \bibinfo{author}{Conti, M.},
  \bibinfo{author}{Bianchi, G.}, \bibinfo{author}{Benedetto, L.},
  \bibinfo{year}{2019}.
\newblock \bibinfo{title}{{A Cyber-Kill-Chain based taxonomy of
  crypto-ransomware features}}.
\newblock \bibinfo{journal}{Journal of Computer Virology and Hacking
  Techniques} \bibinfo{volume}{15}, \bibinfo{pages}{277--305}.
\newblock \URLprefix \url{https://doi.org/10.1007/s11416-019-00338-7},
  \DOIprefix\doi{10.1007/s11416-019-00338-7}.
\bibitem[{Davies et~al.(2021a)Davies, Macfarlane and Buchanan}]{Davies2021}
\bibinfo{author}{Davies, S.R.}, \bibinfo{author}{Macfarlane, R.},
  \bibinfo{author}{Buchanan, W.J.}, \bibinfo{year}{2021}a.
\newblock \bibinfo{title}{{Differential area analysis for ransomware attack
  detection within mixed file datasets}}.
\newblock \bibinfo{journal}{Computers and Security} \bibinfo{volume}{108},
  \bibinfo{pages}{102377}.
\newblock \URLprefix \url{https://doi.org/10.1016/j.cose.2021.102377},
  \DOIprefix\doi{10.1016/j.cose.2021.102377},
  \href{http://arxiv.org/abs/2106.14418}{\tt arXiv:2106.14418}.
\bibitem[{Davies et~al.(2021b)Davies, Macfarlane and Buchanan}]{Davies2021a}
\bibinfo{author}{Davies, S.R.}, \bibinfo{author}{Macfarlane, R.},
  \bibinfo{author}{Buchanan, W.J.}, \bibinfo{year}{2021}b.
\newblock \bibinfo{title}{{NapierOne}}.
\newblock \URLprefix \url{www.napierone.com}.
\bibitem[{Davies et~al.(2022a)Davies, Macfarlane and
  Buchanan}]{davies-entropy-2022}
\bibinfo{author}{Davies, S.R.}, \bibinfo{author}{Macfarlane, R.},
  \bibinfo{author}{Buchanan, W.J.}, \bibinfo{year}{2022}a.
\newblock \bibinfo{title}{{Comparison of Entropy Calculation Methods for
  Ransomware Encrypted File Identification}}.
\newblock \bibinfo{journal}{Entropy} \bibinfo{volume}{24}.
\newblock \DOIprefix\doi{10.3390/e24101503}.
\bibitem[{Davies et~al.(2022b)Davies, Macfarlane and Buchanan}]{Davies2022}
\bibinfo{author}{Davies, S.R.}, \bibinfo{author}{Macfarlane, R.},
  \bibinfo{author}{Buchanan, W.J.}, \bibinfo{year}{2022}b.
\newblock \bibinfo{title}{{NapierOne: A modern mixed file data set alternative
  to Govdocs1}}.
\newblock \bibinfo{journal}{Forensic Science International: Digital
  Investigation} \bibinfo{volume}{40}, \bibinfo{pages}{301330}.
\newblock \URLprefix \url{https://doi.org/10.1016/j.fsidi.2021.301330},
  \DOIprefix\doi{10.1016/j.fsidi.2021.301330}.
\bibitem[{{De Gaspari} et~al.(2020){De Gaspari}, Hitaj, Pagnotta, {De Carli}
  and Mancini}]{DeGaspari2020a}
\bibinfo{author}{{De Gaspari}, F.}, \bibinfo{author}{Hitaj, D.},
  \bibinfo{author}{Pagnotta, G.}, \bibinfo{author}{{De Carli}, L.},
  \bibinfo{author}{Mancini, L.V.}, \bibinfo{year}{2020}.
\newblock \bibinfo{title}{{The Naked Sun: Malicious Cooperation Between
  Benign-Looking Processes}}.
\newblock \bibinfo{journal}{Lecture Notes in Computer Science (including
  subseries Lecture Notes in Artificial Intelligence and Lecture Notes in
  Bioinformatics)} \bibinfo{volume}{12147 LNCS}, \bibinfo{pages}{254--274}.
\newblock \DOIprefix\doi{10.1007/978-3-030-57878-7\_13},
  \href{http://arxiv.org/abs/1911.02423}{\tt arXiv:1911.02423}.
\bibitem[{{De Gaspari} et~al.(2022){De Gaspari}, Hitaj, Pagnotta, {De Carli}
  and Mancini}]{DeGaspari2023}
\bibinfo{author}{{De Gaspari}, F.}, \bibinfo{author}{Hitaj, D.},
  \bibinfo{author}{Pagnotta, G.}, \bibinfo{author}{{De Carli}, L.},
  \bibinfo{author}{Mancini, L.V.}, \bibinfo{year}{2022}.
\newblock \bibinfo{title}{{Evading behavioral classifiers: a comprehensive
  analysis on evading ransomware detection techniques}}.
\newblock \bibinfo{journal}{Neural Computing and Applications}
  \bibinfo{volume}{34}, \bibinfo{pages}{12077--12096}.
\newblock \URLprefix \url{https://doi.org/10.1007/s00521-022-07096-6},
  \DOIprefix\doi{10.1007/s00521-022-07096-6}.
\bibitem[{Dutta et~al.(2022)Dutta, Jadav, Tanwar, Sarma and Pricop}]{Dutta2022}
\bibinfo{author}{Dutta, N.}, \bibinfo{author}{Jadav, N.},
  \bibinfo{author}{Tanwar, S.}, \bibinfo{author}{Sarma, H.K.D.},
  \bibinfo{author}{Pricop, E.}, \bibinfo{year}{2022}.
\newblock \bibinfo{title}{{Introduction to Malware Analysis}}.
  \bibinfo{publisher}{Springer Singapore}, \bibinfo{address}{Singapore}.
  chapter \bibinfo{chapter}{Introducti}.
\newblock pp. \bibinfo{pages}{129--141}.
\newblock \URLprefix \url{https://doi.org/10.1007/978-981-16-6597-4\_7},
  \DOIprefix\doi{10.1007/978-981-16-6597-4\_7}.
\bibitem[{F.R.S.(2009)}]{F.R.S.2009}
\bibinfo{author}{F.R.S., K.P.}, \bibinfo{year}{2009}.
\newblock \bibinfo{title}{{X. On the criterion that a given system of
  deviations from the probable in the case of a correlated system of variables
  is such that it can be reasonably supposed to have arisen from random
  sampling}}.
\newblock \bibinfo{journal}{https://doi.org/10.1080/14786440009463897}
  \bibinfo{volume}{50}, \bibinfo{pages}{157--175}.
\newblock \URLprefix \url{https://www.tandfonline.com/doi/abs/10.1080/
  14786440009463897}, \DOIprefix\doi{10.1080/14786440009463897}.
\bibitem[{Ganfure et~al.(2020)Ganfure, Wu, Chang and Shih}]{Ganfure2020}
\bibinfo{author}{Ganfure, G.O.}, \bibinfo{author}{Wu, C.F.},
  \bibinfo{author}{Chang, Y.H.}, \bibinfo{author}{Shih, W.K.},
  \bibinfo{year}{2020}.
\newblock \bibinfo{title}{{DeepGuard: Deep Generative User-behavior Analytics
  for Ransomware Detection}}.
\newblock \bibinfo{journal}{Proceedings - 2020 IEEE International Conference on
  Intelligence and Security Informatics, ISI 2020}
  \DOIprefix\doi{10.1109/ISI49825.2020.9280508}.
\bibitem[{Gen{\c{c}} et~al.(2018)Gen{\c{c}}, Lenzini and Ryan}]{Genc2018}
\bibinfo{author}{Gen{\c{c}}, Z.A.}, \bibinfo{author}{Lenzini, G.},
  \bibinfo{author}{Ryan, P.Y.}, \bibinfo{year}{2018}.
\newblock \bibinfo{title}{{No random, no ransom: A key to stop cryptographic
  ransomware}}.
\newblock \bibinfo{journal}{Lecture Notes in Computer Science (including
  subseries Lecture Notes in Artificial Intelligence and Lecture Notes in
  Bioinformatics)} \bibinfo{volume}{10885 LNCS}, \bibinfo{pages}{234--255}.
\newblock \DOIprefix\doi{10.1007/978-3-319-93411-2\_11}.
\bibitem[{Gen{\c{c}} et~al.(2019)Gen{\c{c}}, Lenzini and Ryan}]{Genc2019}
\bibinfo{author}{Gen{\c{c}}, Z.A.}, \bibinfo{author}{Lenzini, G.},
  \bibinfo{author}{Ryan, P.Y.A.}, \bibinfo{year}{2019}.
\newblock \bibinfo{title}{{NoCry : No More Secure Encryption Keys for
  Cryptographic Ransomware}}, in: \bibinfo{booktitle}{International Workshop on
  Emerging Technologies for Authorization and Authentication}, pp.
  \bibinfo{pages}{69--85}.
\newblock \DOIprefix\doi{10.1007/978-3-030-39749-4\_5}.
\bibitem[{Gharib and Ghorbani(2017)}]{Gharib2017}
\bibinfo{author}{Gharib, A.}, \bibinfo{author}{Ghorbani, A.},
  \bibinfo{year}{2017}.
\newblock \bibinfo{title}{{DNA-Droid: A real-time android ransomware detection
  framework}}.
\newblock \bibinfo{journal}{Lecture Notes in Computer Science (including
  subseries Lecture Notes in Artificial Intelligence and Lecture Notes in
  Bioinformatics)} \bibinfo{volume}{10394 LNCS}, \bibinfo{pages}{184--198}.
\newblock \DOIprefix\doi{10.1007/978-3-319-64701-2\_14}.
\bibitem[{Google(2015)}]{Google2015}
\bibinfo{author}{Google}, \bibinfo{year}{2015}.
\newblock \bibinfo{title}{{File types indexable by Google}}.
\newblock \URLprefix \url{https://support.google.com/webmasters/answer
  /35287?hl=en}, \URLDATEprefix \urldate{2021-03-24}).
\bibitem[{Halderman et~al.(2009)Halderman, Schoen, Heninger, Clarkson, Paul,
  Calandrino, Feldman, Appelbaum and Felten}]{Halderman2009a}
\bibinfo{author}{Halderman, J.A.}, \bibinfo{author}{Schoen, S.D.},
  \bibinfo{author}{Heninger, N.}, \bibinfo{author}{Clarkson, W.},
  \bibinfo{author}{Paul, W.}, \bibinfo{author}{Calandrino, J.A.},
  \bibinfo{author}{Feldman, A.J.}, \bibinfo{author}{Appelbaum, J.},
  \bibinfo{author}{Felten, E.W.}, \bibinfo{year}{2009}.
\newblock \bibinfo{title}{{Lest we remember: Cold-boot attacks on encryption
  keys}}.
\newblock \bibinfo{journal}{Communications of the ACM} \bibinfo{volume}{52},
  \bibinfo{pages}{91--98}.
\newblock \DOIprefix\doi{10.1145/1506409.1506429}.
\bibitem[{Hall(2006)}]{Hall2006}
\bibinfo{author}{Hall, G.A.}, \bibinfo{year}{2006}.
\newblock \bibinfo{title}{{Sliding Window Measurement for File Type
  Identification}}.
\newblock \bibinfo{journal}{Proceedings of the 1997 ACM symposium on Applied
  computing} , \bibinfo{pages}{529--532}.
\bibitem[{Heninger and Feldman(2008)}]{Heninger2008}
\bibinfo{author}{Heninger, N.}, \bibinfo{author}{Feldman, A.},
  \bibinfo{year}{2008}.
\newblock \bibinfo{title}{{AESKeyFind}}.
\newblock \URLprefix
  \url{https://github.com/eugenekolo/sec-tools/tree/master/crypto/aeskeyfind/aeskeyfind}.
\bibitem[{Homayoun et~al.(2020)Homayoun, Dehghantanha, Ahmadzadeh, Hashemi and
  Khayami}]{Homayoun2020}
\bibinfo{author}{Homayoun, S.}, \bibinfo{author}{Dehghantanha, A.},
  \bibinfo{author}{Ahmadzadeh, M.}, \bibinfo{author}{Hashemi, S.},
  \bibinfo{author}{Khayami, R.}, \bibinfo{year}{2020}.
\newblock \bibinfo{title}{{Know Abnormal, Find Evil: Frequent Pattern Mining
  for Ransomware Threat Hunting and Intelligence}}.
\newblock \bibinfo{journal}{IEEE Transactions on Emerging Topics in Computing}
  \bibinfo{volume}{8}, \bibinfo{pages}{341--351}.
\newblock \DOIprefix\doi{10.1109/TETC.2017.2756908}.
\bibitem[{John et~al.(2022)John, Abbasi, Al-Sahaf and Welch}]{john}
\bibinfo{author}{John, T.C.}, \bibinfo{author}{Abbasi, M.S.},
  \bibinfo{author}{Al-Sahaf, H.}, \bibinfo{author}{Welch, I.},
  \bibinfo{year}{2022}.
\newblock \bibinfo{title}{{Automatically Evolving Malice Scoring Models through
  Utilisation of Genetic Programming: A Cooperative Coevolution Approach}}.
  volume~\bibinfo{volume}{1}.
\newblock \bibinfo{edition}{john} ed., \bibinfo{publisher}{Association for
  Computing Machinery}.
\newblock \DOIprefix\doi{10.1145/3520304.3529063}.
\bibitem[{Joseph and Norman(2020)}]{Joseph2020a}
\bibinfo{author}{Joseph, P.}, \bibinfo{author}{Norman, J.},
  \bibinfo{year}{2020}.
\newblock \bibinfo{title}{{Systematic Memory Forensic Analysis of Ransomware
  using Digital Forensic Tools}}.
\newblock \bibinfo{journal}{International Journal of Natural Computing
  Research} \bibinfo{volume}{9}, \bibinfo{pages}{61--81}.
\newblock \DOIprefix\doi{10.4018/ijncr.2020040105}.
\bibitem[{Kara(2023)}]{KARA2023119133}
\bibinfo{author}{Kara, I.}, \bibinfo{year}{2023}.
\newblock \bibinfo{title}{{Fileless malware threats: Recent advances, analysis
  approach through memory forensics and research challenges}}.
\newblock \bibinfo{journal}{Expert Systems with Applications}
  \bibinfo{volume}{214}, \bibinfo{pages}{119133}.
\newblock \URLprefix \url{https://www.sciencedirect.com/science/article/pii
  /S0957417422021510},
  \DOIprefix\doi{https://doi.org/10.1016/j.eswa.2022.119133}.
\bibitem[{Kessler()}]{garrykessler}
\bibinfo{author}{Kessler, G.}, .
\newblock \bibinfo{title}{{GCK'S File Signature Table}}.
\newblock \URLprefix \url{https://www.garykessler.net/library/file\_sigs.html},
  \URLDATEprefix \urldate{2023-02-10}).
\bibitem[{Kharaz et~al.(2016)Kharaz, Arshad, Mulliner, Robertson, Mulliner and
  Robertson}]{Kharaz2016}
\bibinfo{author}{Kharaz, A.}, \bibinfo{author}{Arshad, S.},
  \bibinfo{author}{Mulliner, C.}, \bibinfo{author}{Robertson, W.},
  \bibinfo{author}{Mulliner, C.}, \bibinfo{author}{Robertson, W.},
  \bibinfo{year}{2016}.
\newblock \bibinfo{title}{{UNVEIL : A Large-Scale , Automated Approach to
  Detecting Ransomware This paper is included in the Proceedings of the}}.
\newblock \bibinfo{journal}{In Proceedings of the 2014 VIRUS BULLETIN
  CONFERENCE} , \bibinfo{pages}{757--772}\URLprefix
  \url{https://www.usenix.org/conference/usenixsecurity16
  /technical-sessions/presentation/kharaz}.
\bibitem[{Kharraz and Kirda(2017a)}]{Kharraz2017a}
\bibinfo{author}{Kharraz, A.}, \bibinfo{author}{Kirda, E.},
  \bibinfo{year}{2017}a.
\newblock \bibinfo{title}{{Redemption: Real-Time Protection Against Ransomware
  at End-Hosts}}.
\newblock \bibinfo{journal}{Lecture Notes in Computer Science (including
  subseries Lecture Notes in Artificial Intelligence and Lecture Notes in
  Bioinformatics)} \bibinfo{volume}{10453 LNCS}, \bibinfo{pages}{98--119}.
\newblock \DOIprefix\doi{10.1007/978-3-319-66332-6\_5}.
\bibitem[{Kharraz and Kirda(2017b)}]{Kharraz2017}
\bibinfo{author}{Kharraz, A.}, \bibinfo{author}{Kirda, E.},
  \bibinfo{year}{2017}b.
\newblock \bibinfo{title}{{Redemption: Real-Time Protection Against Ransomware
  at End-Hosts}}.
\newblock \bibinfo{journal}{Lecture Notes in Computer Science (including
  subseries Lecture Notes in Artificial Intelligence and Lecture Notes in
  Bioinformatics)} \bibinfo{volume}{10453 LNCS}, \bibinfo{pages}{98--119}.
\newblock \DOIprefix\doi{10.1007/978-3-319-66332-6\_5}.
\bibitem[{Ki et~al.(2015)Ki, Kim and Kim}]{Ki2015}
\bibinfo{author}{Ki, Y.}, \bibinfo{author}{Kim, E.}, \bibinfo{author}{Kim,
  H.K.}, \bibinfo{year}{2015}.
\newblock \bibinfo{title}{{A novel approach to detect malware based on API call
  sequence analysis}}.
\newblock \bibinfo{journal}{International Journal of Distributed Sensor
  Networks} \bibinfo{volume}{2015}.
\newblock \URLprefix \url{https://journals.sagepub.com/doi/pdf/10.1155/2015/
  659101}, \DOIprefix\doi{10.1155/2015/659101}.
\bibitem[{Kim et~al.(2022)Kim, Paik, Kim and Cho}]{Kim2022}
\bibinfo{author}{Kim, G.Y.}, \bibinfo{author}{Paik, J.Y.},
  \bibinfo{author}{Kim, Y.}, \bibinfo{author}{Cho, E.S.}, \bibinfo{year}{2022}.
\newblock \bibinfo{title}{{Byte Frequency Based Indicators for
  Crypto-Ransomware Detection from Empirical Analysis}}.
\newblock \bibinfo{journal}{Journal of Computer Science and Technology}
  \bibinfo{volume}{37}, \bibinfo{pages}{423--442}.
\newblock \DOIprefix\doi{10.1007/s11390-021-0263-x}.
\bibitem[{Klein(2006)}]{Klein2006}
\bibinfo{author}{Klein, T.}, \bibinfo{year}{2006}.
\newblock \bibinfo{title}{{All your private keys are belong to us process
  memory}}.
\newblock \bibinfo{journal}{Tech. Rep} , \bibinfo{pages}{1--7}\URLprefix
  \url{https://citeseerx.ist.psu.edu/document?repid=
  rep1\&type=pdf\&doi\newline=cf85042cca0da125b860db7c2fefb38012396cbc}.
\bibitem[{Kornblum(2017)}]{Kornblum}
\bibinfo{author}{Kornblum, J.}, \bibinfo{year}{2017}.
\newblock \bibinfo{title}{findaes}.
\newblock \URLprefix \url{http://jessekornblum.com/tools/findaes/},
  \URLDATEprefix \urldate{2019-08-10}).
\bibitem[{Lebbie et~al.(2022)Lebbie, Prabhu and Agrawal}]{lebbie2022}
\bibinfo{author}{Lebbie, M.}, \bibinfo{author}{Prabhu, S.R.},
  \bibinfo{author}{Agrawal, A.K.}, \bibinfo{year}{2022}.
\newblock \bibinfo{title}{{Comparative Analysis of Dynamic Malware Analysis
  Tools}}, in: \bibinfo{editor}{Dua, M.}, \bibinfo{editor}{Jain, A.K.},
  \bibinfo{editor}{Yadav, A.}, \bibinfo{editor}{Kumar, N.},
  \bibinfo{editor}{Siarry, P.} (Eds.), \bibinfo{booktitle}{Proceedings of the
  International Conference on Paradigms of Communication, Computing and Data
  Sciences}, \bibinfo{publisher}{Springer Singapore},
  \bibinfo{address}{Singapore}. pp. \bibinfo{pages}{359--368}.
\bibitem[{Lee and Lee(2022)}]{lee2022}
\bibinfo{author}{Lee, J.}, \bibinfo{author}{Lee, K.}, \bibinfo{year}{2022}.
\newblock \bibinfo{title}{{A Method for Neutralizing Entropy Measurement-Based
  Ransomware Detection Technologies Using Encoding Algorithms}}.
\newblock \bibinfo{journal}{Entropy} \bibinfo{volume}{24}.
\newblock \DOIprefix\doi{10.3390/e24020239}.
\bibitem[{Lee et~al.(2019a)Lee, Lee and Yim}]{Lee2019a}
\bibinfo{author}{Lee, K.}, \bibinfo{author}{Lee, S.Y.}, \bibinfo{author}{Yim,
  K.}, \bibinfo{year}{2019}a.
\newblock \bibinfo{title}{{Effective Ransomware Detection Using Entropy
  Estimation of Files for Cloud Services}}, in: \bibinfo{booktitle}{I-SPAN
  2019: Pervasive Systems, Algorithms and Networks},
  \bibinfo{publisher}{Springer International Publishing}. pp.
  \bibinfo{pages}{133--139}.
\newblock \URLprefix \url{http://dx.doi.org/10.1007/978-3-030-30143-9\_11},
  \DOIprefix\doi{10.1007/978-3-030-30143-9\_11}.
\bibitem[{Lee et~al.(2019b)Lee, Lee and Yim}]{Lee2019}
\bibinfo{author}{Lee, K.}, \bibinfo{author}{Lee, S.Y.}, \bibinfo{author}{Yim,
  K.}, \bibinfo{year}{2019}b.
\newblock \bibinfo{title}{{Machine Learning Based File Entropy Analysis for
  Ransomware Detection in Backup Systems}}.
\newblock \bibinfo{journal}{IEEE Access} \bibinfo{volume}{7},
  \bibinfo{pages}{110205--110215}.
\newblock \URLprefix \url{https://dx.doi.org/10.1109/ACCESS.2019.2931136},
  \DOIprefix\doi{10.1109/access.2019.2931136}.
\bibitem[{Lemmou et~al.(2021)Lemmou, Lanet and Souidi}]{Lemmou2021c}
\bibinfo{author}{Lemmou, Y.}, \bibinfo{author}{Lanet, J.L.},
  \bibinfo{author}{Souidi, E.M.}, \bibinfo{year}{2021}.
\newblock \bibinfo{title}{{In-depth analysis of ransom note files}}.
\newblock \bibinfo{journal}{Computers} \bibinfo{volume}{10},
  \bibinfo{pages}{1--25}.
\newblock \DOIprefix\doi{10.3390/computers10110145}.
\bibitem[{Leommoore()}]{leommoore}
\bibinfo{author}{Leommoore}, .
\newblock \bibinfo{title}{{File Magic Numbers {\textperiodcentered} GitHub}}.
\newblock \URLprefix
  \url{https://gist.github.com/leommoore/f9e57ba2aa4bf197ebc5}, \URLDATEprefix
  \urldate{2023-02-28}).
\bibitem[{Li et~al.(2005)Li, Wang, Stolfo and Herzog}]{Li2005}
\bibinfo{author}{Li, W.J.}, \bibinfo{author}{Wang, K.},
  \bibinfo{author}{Stolfo, S.J.}, \bibinfo{author}{Herzog, B.},
  \bibinfo{year}{2005}.
\newblock \bibinfo{title}{{Fileprints: Identifying file types by n-gram
  analysis}}.
\newblock \bibinfo{journal}{Proceedings from the 6th Annual IEEE System, Man
  and Cybernetics Information Assurance Workshop, SMC 2005}
  \bibinfo{volume}{2005}, \bibinfo{pages}{64--71}.
\newblock \URLprefix \url{https://dx.doi.org/10.1109/IAW.2005.1495935},
  \DOIprefix\doi{10.1109/IAW.2005.1495935}.
\bibitem[{Lokuketagoda et~al.(2018)Lokuketagoda, Weerakoon, Kuruppu, Senarathne
  and {Yapa Abeywardena}}]{Lokuketagoda2018}
\bibinfo{author}{Lokuketagoda, B.}, \bibinfo{author}{Weerakoon, M.P.},
  \bibinfo{author}{Kuruppu, U.M.}, \bibinfo{author}{Senarathne, A.N.},
  \bibinfo{author}{{Yapa Abeywardena}, K.}, \bibinfo{year}{2018}.
\newblock \bibinfo{title}{{R - Killer: An email based ransomware protection
  tool}}.
\newblock \bibinfo{journal}{13th International Conference on Computer Science
  and Education, ICCSE 2018} ,
  \bibinfo{pages}{735--741}\DOIprefix\doi{10.1109/ICCSE.2018.8468807}.
\bibitem[{Maartmann-Moe et~al.(2009)Maartmann-Moe, Thorkildsen and
  {\AA}rnes}]{Maartmann-Moe2009b}
\bibinfo{author}{Maartmann-Moe, C.}, \bibinfo{author}{Thorkildsen, S.E.},
  \bibinfo{author}{{\AA}rnes, A.}, \bibinfo{year}{2009}.
\newblock \bibinfo{title}{{The persistence of memory: Forensic identification
  and extraction of cryptographic keys}}.
\newblock \bibinfo{journal}{DFRWS 2009 Annual Conference} \bibinfo{volume}{6},
  \bibinfo{pages}{132--140}.
\newblock \DOIprefix\doi{10.1016/j.diin.2009.06.002}.
\bibitem[{Maigida et~al.(2019)Maigida, Abdulhamid, Olalere, Alhassan, Chiroma
  and Dada}]{Maigida2019}
\bibinfo{author}{Maigida, A.M.}, \bibinfo{author}{Abdulhamid, S.M.},
  \bibinfo{author}{Olalere, M.}, \bibinfo{author}{Alhassan, J.K.},
  \bibinfo{author}{Chiroma, H.}, \bibinfo{author}{Dada, E.G.},
  \bibinfo{year}{2019}.
\newblock \bibinfo{title}{{Systematic literature review and metadata analysis
  of ransomware attacks and detection mechanisms}}.
\newblock \bibinfo{journal}{Journal of Reliable Intelligent Environments}
  \bibinfo{volume}{5}, \bibinfo{pages}{67--89}.
\newblock \URLprefix \url{https://doi.org/10.1007/s40860-019-00080-3},
  \DOIprefix\doi{10.1007/s40860-019-00080-3}.
\bibitem[{MalwareBytes(2023)}]{ION}
\bibinfo{author}{MalwareBytes}, \bibinfo{year}{2023}.
\newblock \bibinfo{title}{{ION starts bringing customers back online after
  LockBit ransomware attack}}.
\newblock \URLprefix
  \url{https://www.malwarebytes.com/blog/news/2023/02/ion-starts-bringing-customers-back-online-after-lockbit-ransomware-attack},
  \URLDATEprefix \urldate{2023-02-20}).
\bibitem[{Manavi and Hamzeh(2022)}]{Manavi2022}
\bibinfo{author}{Manavi, F.}, \bibinfo{author}{Hamzeh, A.},
  \bibinfo{year}{2022}.
\newblock \bibinfo{title}{{A novel approach for ransomware detection based on
  PE header using graph embedding}}.
\newblock \bibinfo{journal}{Journal of Computer Virology and Hacking
  Techniques} \URLprefix \url{https://doi.org/10.1007/s11416-021-00414-x},
  \DOIprefix\doi{10.1007/s11416-021-00414-x}.
\bibitem[{McDonald et~al.(2022)McDonald, Papadopoulos, Pitropakis, Ahmad and
  Buchanan}]{McDonald2022}
\bibinfo{author}{McDonald, G.}, \bibinfo{author}{Papadopoulos, P.},
  \bibinfo{author}{Pitropakis, N.}, \bibinfo{author}{Ahmad, J.},
  \bibinfo{author}{Buchanan, W.J.}, \bibinfo{year}{2022}.
\newblock \bibinfo{title}{{Ransomware: Analysing the Impact on Windows Active
  Directory Domain Services}}.
\newblock \bibinfo{journal}{Sensors} \bibinfo{volume}{22}.
\newblock \DOIprefix\doi{10.3390/s22030953},
  \href{http://arxiv.org/abs/2202.03276}{\tt arXiv:2202.03276}.
\bibitem[{McIntosh et~al.(2019)McIntosh, Jang-Jaccard, Watters and
  Susnjak}]{mcintosh}
\bibinfo{author}{McIntosh, T.}, \bibinfo{author}{Jang-Jaccard, J.},
  \bibinfo{author}{Watters, P.}, \bibinfo{author}{Susnjak, T.},
  \bibinfo{year}{2019}.
\newblock \bibinfo{title}{{The Inadequacy of Entropy-Based Ransomware
  Detection}}, in: \bibinfo{booktitle}{International Conference on Neural
  Information Processing}, \bibinfo{publisher}{Springer International
  Publishing}. pp. \bibinfo{pages}{181--189}.
\newblock \URLprefix \url{https://dx.doi.org/10.1007/978-3-030-36802-9\_20},
  \DOIprefix\doi{10.1007/978-3-030-36802-9\_20}.
\bibitem[{Mehnaz et~al.(2018)Mehnaz, Mudgerikar and Bertino}]{Mehnaz2018}
\bibinfo{author}{Mehnaz, S.}, \bibinfo{author}{Mudgerikar, A.},
  \bibinfo{author}{Bertino, E.}, \bibinfo{year}{2018}.
\newblock \bibinfo{title}{{RWGuard: A Real-Time Detection System Against
  Cryptographic Ransomware}}, in: \bibinfo{booktitle}{International Symposium
  on Research in Attacks, Intrusions, and Defenses}.
  \bibinfo{publisher}{Springer International Publishing}.
  volume~\bibinfo{volume}{1}, pp. \bibinfo{pages}{114--136}.
\newblock \URLprefix \url{https://doi.org/10.1007/978-3-030-00470-5\_6},
  \DOIprefix\doi{10.1007/978-3-030-00470-5\_6}.
\bibitem[{Moser et~al.(2007)Moser, Kruegel and Kirda}]{moser}
\bibinfo{author}{Moser, A.}, \bibinfo{author}{Kruegel, C.},
  \bibinfo{author}{Kirda, E.}, \bibinfo{year}{2007}.
\newblock \bibinfo{title}{{Limits of Static Analysis for Malware Detection}},
  in: \bibinfo{booktitle}{Twenty-Third Annual Computer Security Applications
  Conference (ACSAC 2007)}, pp. \bibinfo{pages}{421--430}.
\newblock \DOIprefix\doi{10.1109/ACSAC.2007.21}.
\bibitem[{Nieuwenhuizen(2017)}]{Nieuwenhuizen}
\bibinfo{author}{Nieuwenhuizen, D.}, \bibinfo{year}{2017}.
\newblock \bibinfo{title}{{A behavioural-based approach to ransomware
  detection}}.
\newblock \bibinfo{journal}{Computer Science} \URLprefix
  \url{https://www.semanticscholar.org/paper/A-behavioural-
  based-approach-to-ransomware-Nieuwenhuizen
  93b6e2fdf2a79608e44ab64e37bddd6973f54f1d}.
\bibitem[{O'Kane et~al.(2011)O'Kane, Sezer and McLaughlin}]{kane}
\bibinfo{author}{O'Kane, P.}, \bibinfo{author}{Sezer, S.},
  \bibinfo{author}{McLaughlin, K.}, \bibinfo{year}{2011}.
\newblock \bibinfo{title}{{Obfuscation: The Hidden Malware}}.
\newblock \bibinfo{journal}{IEEE Security \& Privacy} \bibinfo{volume}{9},
  \bibinfo{pages}{41--47}.
\newblock \DOIprefix\doi{10.1109/MSP.2011.98}.
\bibitem[{Oz et~al.(2021)Oz, Aris, Levi and Uluagac}]{Oz2021}
\bibinfo{author}{Oz, H.}, \bibinfo{author}{Aris, A.}, \bibinfo{author}{Levi,
  A.}, \bibinfo{author}{Uluagac, A.S.}, \bibinfo{year}{2021}.
\newblock \bibinfo{title}{{A Survey on Ransomware: Evolution, Taxonomy, and
  Defense Solutions}}.
\newblock \bibinfo{journal}{ACM Computing Surveys} \bibinfo{volume}{1}.
\newblock \URLprefix \url{http://arxiv.org/abs/2102.06249},
  \href{http://arxiv.org/abs/2102.06249}{\tt arXiv:2102.06249}.
\bibitem[{{Pattern Recognition and Applications Lab}(2023)}]{packdroid}
\bibinfo{author}{{Pattern Recognition and Applications Lab}},
  \bibinfo{year}{2023}.
\newblock \bibinfo{title}{{R-PackDroid Ransomware Detector (App) and Dataset |
  PRA Lab}}.
\newblock \URLprefix \url{http://pralab.diee.unica.it/en/RPackDroid},
  \URLDATEprefix \urldate{2023-02-23}).
\bibitem[{Prachi and Kumar(2022)}]{Prachi2022}
\bibinfo{author}{Prachi}, \bibinfo{author}{Kumar, S.}, \bibinfo{year}{2022}.
\newblock \bibinfo{title}{{An effective ransomware detection approach in a
  cloud environment using volatile memory features}}.
\newblock \bibinfo{journal}{Journal of Computer Virology and Hacking
  Techniques} \URLprefix \url{https://doi.org/10.1007/s11416-022-00425-2},
  \DOIprefix\doi{10.1007/s11416-022-00425-2}.
\bibitem[{Sai and Kumar(2019)}]{Sai2019}
\bibinfo{author}{Sai, R.L.P.}, \bibinfo{author}{Kumar, T.P.},
  \bibinfo{year}{2019}.
\newblock \bibinfo{title}{{Reverse Engineering the Behaviour of NotPetya
  Ransomware}}.
\newblock \bibinfo{journal}{International Journal of Recent Technology and
  Engineering} , \bibinfo{pages}{574--578}\URLprefix
  \url{https://www.ijrte.org/wp-content/uploads/papers/v7i6s/F03120376S19.pdf}.
\bibitem[{Salehi et~al.(2018a)Salehi, Shahriari, Ahmadian and
  Tazik}]{Salehi2018}
\bibinfo{author}{Salehi, S.}, \bibinfo{author}{Shahriari, H.},
  \bibinfo{author}{Ahmadian, M.M.}, \bibinfo{author}{Tazik, L.},
  \bibinfo{year}{2018}a.
\newblock \bibinfo{title}{{A Novel Approach for Detecting DGA-based
  Ransomwares}}.
\newblock \bibinfo{journal}{2018 15th International ISC (Iranian Society of
  Cryptology) Conference on Information Security and Cryptology, ISCISC 2018}
  \DOIprefix\doi{10.1109/ISCISC.2018.8546941}.
\bibitem[{Salehi et~al.(2018b)Salehi, Shahriari, Ahmadian and Tazik}]{salehi}
\bibinfo{author}{Salehi, S.}, \bibinfo{author}{Shahriari, H.},
  \bibinfo{author}{Ahmadian, M.M.}, \bibinfo{author}{Tazik, L.},
  \bibinfo{year}{2018}b.
\newblock \bibinfo{title}{{A Novel Approach for Detecting DGA-based
  Ransomwares}}, in: \bibinfo{booktitle}{2018 15th International ISC (Iranian
  Society of Cryptology) Conference on Information Security and Cryptology
  (ISCISC)}, pp. \bibinfo{pages}{1--7}.
\newblock \DOIprefix\doi{10.1109/ISCISC.2018.8546941}.
\bibitem[{Scaife et~al.(2016a)Scaife, Carter, Traynor and Butler}]{Scaife2016}
\bibinfo{author}{Scaife, N.}, \bibinfo{author}{Carter, H.},
  \bibinfo{author}{Traynor, P.}, \bibinfo{author}{Butler, K.R.},
  \bibinfo{year}{2016}a.
\newblock \bibinfo{title}{{CryptoLock (and Drop It): Stopping Ransomware
  Attacks on User Data}}.
\newblock \bibinfo{journal}{2016 IEEE 36th International Conference on
  Distributed Computing Systems (ICDCS)} , \bibinfo{pages}{303--312}\URLprefix
  \url{https://doi.org/10.1109/ICDCS.2016.46},
  \DOIprefix\doi{10.1109/ICDCS.2016.46}.
\bibitem[{Scaife et~al.(2016b)Scaife, Carter, Traynor and Butler}]{Scaife2016a}
\bibinfo{author}{Scaife, N.}, \bibinfo{author}{Carter, H.},
  \bibinfo{author}{Traynor, P.}, \bibinfo{author}{Butler, K.R.},
  \bibinfo{year}{2016}b.
\newblock \bibinfo{title}{{CryptoLock (and Drop It): Stopping Ransomware
  Attacks on User Data}}.
\newblock \bibinfo{journal}{Proceedings - International Conference on
  Distributed Computing Systems} \bibinfo{volume}{2016-Augus},
  \bibinfo{pages}{303--312}.
\newblock \DOIprefix\doi{10.1109/ICDCS.2016.46}.
\bibitem[{Shannon(1948)}]{Shannon1948}
\bibinfo{author}{Shannon, C.}, \bibinfo{year}{1948}.
\newblock \bibinfo{title}{{A Mathematical Theory of Communication}}.
\newblock \bibinfo{journal}{Bell System Technology} \bibinfo{volume}{27},
  \bibinfo{pages}{379--423}.
\newblock \URLprefix
  \url{https://dx.doi.org/10.1002/j.1538-7305.1948.tb01338.x},
  \DOIprefix\doi{10.1002/j.1538-7305.1948.tb01338.x}.
\bibitem[{Sheen et~al.(2022)Sheen, Asmitha and Venkatesan}]{Sheen2022}
\bibinfo{author}{Sheen, S.}, \bibinfo{author}{Asmitha, K.A.},
  \bibinfo{author}{Venkatesan, S.}, \bibinfo{year}{2022}.
\newblock \bibinfo{title}{{R-Sentry: Deception based ransomware detection using
  file access patterns}}.
\newblock \bibinfo{journal}{Computers and Electrical Engineering}
  \bibinfo{volume}{103}, \bibinfo{pages}{108346}.
\newblock \URLprefix \url{https://doi.org/10.1016/j.compeleceng.2022.108346},
  \DOIprefix\doi{10.1016/j.compeleceng.2022.108346}.
\bibitem[{Song et~al.(2016)Song, Kim and Lee}]{Song2016}
\bibinfo{author}{Song, S.}, \bibinfo{author}{Kim, B.}, \bibinfo{author}{Lee,
  S.}, \bibinfo{year}{2016}.
\newblock \bibinfo{title}{{The Effective Ransomware Prevention Technique Using
  Process Monitoring on Android Platform}}.
\newblock \bibinfo{journal}{Mobile Information Systems} \bibinfo{volume}{2016},
  \bibinfo{pages}{2946735}.
\newblock \URLprefix \url{https://doi.org/10.1155/2016/2946735},
  \DOIprefix\doi{10.1155/2016/2946735}.
\bibitem[{{The Telegraph Media Group}(2023)}]{royalmail}
\bibinfo{author}{{The Telegraph Media Group}}, \bibinfo{year}{2023}.
\newblock \bibinfo{title}{{Royal Mail turned down {\pounds}66m ransom demand
  from Lockbit hackers}}.
\newblock \URLprefix
  \url{https://www.telegraph.co.uk/business/2023/02/14/royal-mail-turned-66m-ransom-demand-lockbit-hackers/}.
\bibitem[{VandenBrink(2016)}]{VandenBrink2016}
\bibinfo{author}{VandenBrink, R.}, \bibinfo{year}{2016}.
\newblock \bibinfo{title}{{Using File Entropy to Identify "Ransomwared"
  Files}}.
\newblock \URLprefix \url{https://isc.sans.edu/forums/diary/Using+File
  Entropy+to+Identify+Ransomwared+Files/21351/}, \URLDATEprefix
  \urldate{2020-09-05}).
\bibitem[{Walker(2008)}]{walker2008}
\bibinfo{author}{Walker, J.}, \bibinfo{year}{2008}.
\newblock \bibinfo{title}{{Pseudorandom Number Sequence Test Program}}.
\newblock \URLprefix \url{https://www.fourmilab.ch/random/}, \URLDATEprefix
  \urldate{2022-05-29}).
\bibitem[{Wikipedia(a)}]{wikimagicnumber}
\bibinfo{author}{Wikipedia}, a.
\newblock \bibinfo{title}{{List of file formats}}.
\newblock \URLprefix
  \url{https://en.wikipedia.org/wiki/List\_of\_file\_formats}, \URLDATEprefix
  \urldate{2022-01-11}).
\bibitem[{Wikipedia(b)}]{wikilistoffilesignatures}
\bibinfo{author}{Wikipedia}, b.
\newblock \bibinfo{title}{{List of file signatures}}.
\newblock \URLprefix
  \url{https://en.wikipedia.org/wiki/List\_of\_file\_signatures},
  \URLDATEprefix \urldate{2022-01-20}).
\bibitem[{Yamany et~al.(2022)Yamany, Elsayed, Jurcut, Abdelbaki and
  Azer}]{Yamany2022}
\bibinfo{author}{Yamany, B.}, \bibinfo{author}{Elsayed, M.S.},
  \bibinfo{author}{Jurcut, A.D.}, \bibinfo{author}{Abdelbaki, N.},
  \bibinfo{author}{Azer, M.A.}, \bibinfo{year}{2022}.
\newblock \bibinfo{title}{{A New Scheme for Ransomware Classification and
  Clustering Using Static Features}}.
\newblock \bibinfo{journal}{Electronics (Switzerland)} \bibinfo{volume}{11}.
\newblock \DOIprefix\doi{10.3390/electronics11203307}.

\end{thebibliography}
\newpage
\begin{appendices}
\label{appendix}
\section{Ransomware Strains}
\begin{tiny}
\begin{table*}[hp!]
\setlength{\tabcolsep}{4pt}
\caption{SHA256 Hashes of Ransomware Strains Used}
\centering
\begin{tabular}{lll}
\toprule
\textbf{Strain} & \textbf{}  & \textbf{SHA256 Hash} \\
\midrule
AVOSLOCKER   &  & 718810b8eeb682fc70df602d952c0c83e028c5a5bfa44c506756980caf2edebb \\
BADRABBIT    &  & 630325cac09ac3fab908f903e3b00d0dadd5fdaa0875ed8496fcbb97a558d0da \\
BLACKBASTA   &  & 5d2204f3a20e163120f52a2e3595db19890050b2faa96c6cba6b094b0a52b0aa \\
BLACKCAT     &  & 847fb7609f53ed334d5affbb07256c21cb5e6f68b1cc14004f5502d714d2a456 \\
BLACKMATTER  &  & be5bc29f58b868f4ff8cd66b4526535593e515a697bb8951c625bdfed13cccb7 \\
CERBER       &  & e67834d1e8b38ec5864cfa101b140aeaba8f1900a6e269e6a94c90fcbfe56678 \\
CHIMERA      &  & 1dacdc296fd6ef6ba817b184cce9901901c47c01d849adfa4222bfabfed61838 \\
CLOP         &  & a867deb1578088d066941c40e598e4523ab5fd6c3327d3afb951073bee59fb02 \\
CONTI        &  & 2fc6d7df9252b1e2c4eb3ad7d0d29c188d87548127c44cebc40db9abe8e5aa35 \\
CRYPTOLOCKER &  & 5e902a138174c34e5445685c82b2044e0b35565854471aaccef0315c77288dc9 \\
CUBA         &  & 936119bc1811aeef01299a0150141787865a0dbe2667288f018ad24db5a7bc27 \\
DARKSIDE     &  & 508dd6f7ed6c143cf5e1ed6a4051dd8ee7b5bf4b7f55e0704d21ba785f2d5add \\
DHARMA       &  & c2ab289cbd2573572c39cac3f234d77fdf769e48a1715a14feddaea8ae9d9702 \\
GANDCRAB     &  & 64d341ecbc52f9d78080bf23559ec1778824979dd19498ee44032ec1d5224ff6 \\
HELLOKITTY   &  & 501487b025f25ddf1ca32deb57a2b4db43ccf6635c1edc74b9cff54ce0e5bcfe \\
JIGSAW       &  & 3d8d58e8b7431c871121c2d3669f68054eda11cbfb735a3ab689734d4544dab3 \\
LOCKBIT      &  & 43ced481e0f68fe57be3246cc5aede353c9d34f4e15d0afe443b5de9514d3ce4 \\
LORENZ       &  & 1264b40feaa824d5ba31cef3c8a4ede230c61ef71c8a7994875deefe32bd8b3d \\
MAZE         &  & 3885589a3c94d0475a6d994e4644e682f4cff93f8b4d65f37508ffe706861363 \\
MEDUSALOCKER &  & 2e9fceb91d4378a4e67250f0cb633a020be6eb1c57237272a50cb4db36997db7 \\
NETWALKER    &  & 57cf4470348e3b5da0fa3152be84a81a5e2ce5d794976387be290f528fa419fd \\
NOTPETYA     &  & b53f3c0cd32d7f20849850768da6431e5f876b7bfa61db0aa0700b02873393fa \\
PHOBOS       &  & a91491f45b851a07f91ba5a200967921bf796d38677786de51a4a8fe5ddeafd2 \\
RAGNAR       &  & 5469182495d92a5718e0e1dcdf371e92b79724e427050154f318de693d341c89 \\
RANSOMEXX    &  & 4cae449450c07b7aa74314173c7b00d409eabfe22b86859f3b3acedd66010458 \\
RYUK         &  & d083ecc1195602c45d9cb75a08c395ad7d2b0bf73d7e70e2fc76101c780dd38f \\
SODINOKIBI   &  & 06b323e0b626dc4f051596a39f52c46b35f88ea6f85a56de0fd76ec73c7f3851 \\
SUNCRYPT     &  & 0d7ed584dd1ae3cc071ad1b2400a5c534d19206be7a98a6046959a7267c063a1 \\
TESLACRYPT   &  & 4de6675c089aad8a52993b1a21afd06dc7086f4ea948755c09a7a8471e4fddbd \\
WANNACRY     &  & 32f24601153be0885f11d62e0a8a2f0280a2034fc981d8184180c5d3b1b9e8cf \\
WASTEDLOCKER &  & 905ea119ad8d3e54cd228c458a1b5681abc1f35df782977a23812ec4efa0288a \\
\bottomrule
\end{tabular}
\label{tab:ransomwarehash}
\end{table*}
\end{tiny}

\section{Program Information}
\begin{tiny}
\begin{table*}[hp!]
\setlength{\tabcolsep}{4pt}
\caption{Details of Benign Programs Used}
\centering
\begin{tabular}{lllll}
\toprule
\textbf{Program Name} & \textbf{}  & \textbf{File Type} & \textbf{}  & \textbf{Version} \\
\midrule

7z.exe	              & & 7ZIP  & &	9.2  \\
Apk Installer on WSA  & & APK   & &	Apk Installer 4.7 \\
mspaint.exe	          & & BMP   & &	Version 20H2 19042.1466\\
Visual Studio Code	  & & CSS	& & Visual Studio Code v 1.73.1 \\
CSVed.exe	          & & CSV	& & Csved 2.5.6 \\
devenv.exe	          & & DLL	& & Version 4.8.04084 \\
wordpad.exe	          & & DOC	& & Version 20H2 19042.1466\\
winword.exe	          & & DOCX	& & Version 15.0.4430.1017 \\
DWGFastview.exe(gcad) & & DWG	& & build 221110 -  32bit V6.0.0\\
cl.exe	              & & ELF	& & Version  19.29.30139 \\
scribus.exe	          & & EPS	& & Version 1.4.8\\
msEdge.exe	          & & EPUB	& & Version 106.0.1370.52 \\
explorer.exe	      & & EXE	& & 22H2 (10.0.22621.675) \\
MicrosoftPhoto.exe	  & & GIF	& & Microsoft Photos 2022.30070.26007.0 \\
gzip.exe	          & & GZIP	& & Version 1.3.12 \\
firefox.exe	          & & HTML	& & 105.0.3  \\
outlook.exe	          & & ICS	& & Version 15.0.4430.1017 \\
eclipse.exe	          & & JS	& & Version: 2020-03 (4.15.0) uild id: 20200313-1211\\
3DViewer.exe	      & & JPG	& & 7.2107.7012.0 \\
Altova XMLSpy	      & & JSON	& & Version 2023 \\
wmplayer.exe	      & & MKV	& & 12.0.22621.457 \\
Spotify.exe	          & & MP3	& & 1.1.95.893.g6cf4d40c\\
VLC.exe	              & & MP4	& & Version 3.0.16\\
soffice.exe	          & & ODS	& & 7.4.3.3\\
XPSViewer	          & & OXPS	& & Version 20H2 19042.1466\\
AcroRd32.exe	      & & PDF	& & Version 22.768\\
GIMP	              & & PNG	& & Version 2.10\\
powershell\_ise.exe	  & & PS	& & 5.1.19041.1320\\
WPS Office (wps.exe)  & & PPT	& & 11.2.0.11388\\ 
PowerPoint.exe	      & & PPTX	& & Version 15.0.4430.1017\\
WinRAR.exe	          & & RAR	& & 3.42\\
Chrome.exe	          & & SVG	& & Version 106.0.5249.119\\
tar.exe	              & & TAR	& & bsdtar 3.5.2 - libarchive 3.5.2 zlib/1.2.5.f-ipp \\
photoshop.exe	      & & TIF	& & 13.1-2x32\\
Notepad.exe	          & & TXT	& & Version 20H2 19042.1466\\
iexplorer.exe	      & & WEBP	& & 11.630.19041.0\\
OpenOffice.exe	      & & XLS	& & 4.1.3\\
Excel.exe	          & & XLSX	& & Version 15.0.4430.1017\\
notepad++.exe	      & & XML	& & 7.9.1\\
winzip64.exe	      & & ZIP	& & 27.0 15240\\
minigz.exe	          & & ZLIB & & Version 1	\\
\bottomrule
\end{tabular}
\label{tab:benignprograms}
\end{table*}
\end{tiny}

\end{appendices}

\end{document}